\documentclass[fleqn,usenatbib]{mnras}

\usepackage[T1]{fontenc}

\DeclareRobustCommand{\VAN}[3]{#2}
\let\VANthebibliography\thebibliography
\def\thebibliography{\DeclareRobustCommand{\VAN}[3]{##3}\VANthebibliography}

\usepackage{graphicx}
\usepackage{float}
\usepackage{amsmath}	

\graphicspath{{img/}}

\usepackage{color}

\newcommand{\cm}{cm$^{-1}$}

\newcommand{\bra}[1]{\langle #1|}
\newcommand{\ket}[1]{|#1\rangle}

\newcommand{\pgopher}{\textsc{Pgopher}}
\newcommand{\exocross}{\textsc{ExoCross}}
\newcommand{\molpro}{\textsc{Molpro}}

\newcommand{\duo}{{\sc Duo}}
\newcommand{\Duo}{{\sc Duo}}

\newcommand{\ai}{\textit{ab initio}}

\newcommand{\XS}{$X\,{}^{2}\Sigma^{+}$}
\newcommand{\AS}{$A\,{}^{2}\Pi$}
\newcommand{\CS}{$C\,{}^{2}\Pi$}
\newcommand{\ApS}{$A'\,{}^{2}\Delta$}
\newcommand{\BS}{$B\,{}^{2}\Sigma^{+}$}
\newcommand{\DS}{$D\,{}^{2}\Sigma^{+}$}

\newcommand{\XSM}{X\,{}^{2}\Sigma^{+}}
\newcommand{\ASM}{A\,{}^{2}\Pi}
\newcommand{\CSM}{C\,{}^{2}\Pi}

\newcommand{\BSM}{B\,{}^{2}\Sigma^{+}}
\newcommand{\DSM}{D\,{}^{2}\Sigma^{+}}

\newcommand{\band}[2]{#1$\xrightarrow[]{}$#2}

\newcommand{\name}{BRYTS}

\title[ExoMol line lists -- {LIII}. YO]{ExoMol line lists -- LIII: Empirical Rovibronic spectra of Yttrium Oxide (YO)}

\author[Yurchenko et al.]{
Sergei N. Yurchenko,$^{1}$
Ryan P. Brady,$^{1}$
Jonathan Tennyson,$^{1}$\thanks{The corresponding author: j.tennyson@ucl.ac.uk}
Alexander N. Smirnov,$^{2}$
\newauthor 
Oleg A. Vasilyev$^{3}$
and Victor G. Solomonik$^{2}$
\\
$^1$ Department of Physics and Astronomy, University College London, Gower Street, WC1E 6BT London, UK\\
$^2$ Department of Physics, Ivanovo State University of Chemistry and Technology, Sheremetev Ave., 7, Ivanovo, 153000, Russia\\
$^3$ Department of Chemistry and Biochemistry, The Ohio State University, Columbus, OH 43210, U.S.A.
}

\date{Accepted XXXX. Received XXXX; in original form XXXX}

\date{\today}

\begin{document}

\label{firstpage}

\maketitle

\pagerange{\pageref{firstpage}--\pageref{lastpage}}

\begin{abstract}

Empirical line lists BRYTS for the open shell  molecule $^{89}$Y$^{16}$O (yttrium oxide) and its isotopologues are presented. The  line lists cover the 6 lowest electronic states: $X\,{}^{2}\Sigma^{+}$, $A\,{}^{2}\Pi$, $A'\,{}^{2}\Delta$, $B\,{}^{2}\Sigma^{+}$, $C\,{}^{2}\Pi$ and $D\,{}^{2}\Sigma^{+}$  up to 60~000~cm$^{-1}$ ($<0.167$~$\mu$m) for rotational excitation up to  $J = 400.5$. An  \textit{ab initio} spectroscopic  model consisting of potential energy curves (PECs), spin-orbit and electronic angular momentum couplings is refined by fitting to experimentally determined energies of YO, derived from published YO experimental transition frequency data. The model is complemented by empirical spin-rotation and $\Lambda$-doubling curves and \ai\ dipole moment and transition dipole moment curves computed using MRCI. The \ai\ PECs computed using the  complete basis set limit extrapolation and the CCSD(T) method with its higher quality  provide an excellent initial approximation for the refinement.  Non-adiabatic coupling curves for two pairs of states  of the same symmetry $A$/$C$ and $B$/$D$  are computed using a state-averaged CASSCF and used to build diabatic representations for the $A\,{}^{2}\Pi$, $C\,{}^{2}\Pi$, $B\,{}^{2}\Sigma^{+}$ and  $D\,{}^{2}\Sigma^{+}$ curves. The experimentally derived energies of $^{89}$Y$^{16}$O utilised in the fit are used to replace the corresponding calculated energy values in the BRYTS line list. Simulated spectra of YO show excellent agreement with the experiment, where it is available. Calculated lifetimes of YO are tuned to agree well with the experiment, where available. The BRYTS YO line lists are available from the ExoMol database  (\url{www.exomol.com}).

\end{abstract}

\begin{keywords}
molecular data - exoplanets - stars: atmospheres - stars: low-mass
\end{keywords}



\section{Introduction}

The spectrum of yttrium oxide, YO, has been the subject of many astrophysical studies. It has been  observed in the spectra of cool stars \citep{78wYcL.YO} including R-Cygni \citep{78Sauval,82Murty.YO}, Pi-Gruis \citep{83Murty.YO}, V838 Mon \citep{07GoBaxx.YO,09KaScTy.YO}, and V4332 Sgr \citep{07GoBaxx.YO,15TyGoKa}. YO has also been actively used in laser cooling  experiments \citep{15YeHuCo.YO,15CoHuYe.YO,16QuGoJo.YO,18CoDiWu.YO}. Its spectrum has been used as a probe to study high-temperature materials \citep{05BaCaG1.YO}.

There are many laboratory spectroscopic studies of YO, including its  \AS\ -- \XS\
\citep{77ShNixx.YO,78Linton.YO,79BeBaLu.YO,79LiPaxx.YO,80WiDiZe.YO,82BaMuxx.YO,83BeGrxx.YO,84WiDiZe.YO,88ChPoSt.YO,90StShxx.YO,91DyMuNo.YO,93FrKuRe.YO,93OtGoxx.YO,02BaGrxx.YO,03BaGrxx.YO,05BaCaG1.YO,05BaCaGr.YO,06KoSexx.YO,07BaCaG1.YO,07BaCaGr.YO,15CoHuYe.YO,23MuNa.YO},
\BS\ -- \XS\ \citep{77ShNixx.YO,79BeBaLu.YO,80BeGrxx.YO,93FrKuRe.YO,05LeMaCh.YO,17ZhZhZh.YO}, \ApS\ -- \XS
\citep{76ChGoxx.YO,92SiJaHa.YO,15CoHuYe.YO} and \DS\ -- \XS\ \citep{17ZhZhZh.YO}  band systems,  rotational spectrum \citep{61UhAkxx.YO,86StAlxx.YO,93HoToxx.YO}, hyperfine spectrum \citep{65KaWexx.YO,86StAlxx.YO,87StAlxx.YO,88ChPoSt.YO,90SuLoFr.YO,99KnKaPe.YO,03StVixx.YO} and
chemiluminescence spectra \citep{75MaPaxx.YO,77ChGoxx.YO,93FrKuRe.YO}.  The very recent experimental study of the \AS\ and \BS\ systems of YO by \citet{23MuNa.YO} provided  crucial information for this work on the coupling between the \BS\ and \DS\ states.  Relative intensity measurements of  the \AS\ -- \XS\ system were performed by \citet{82BaMuxx.YO}. Permanent dipole moments of YO in both the \XS\ and \AS\ states have been measured using the Stark technique  \citep{90StShxx.YO,90SuLoFr.YO,03StVixx.YO}.
The lifetimes in the \AS, \BS, and \DS\ states were measured by \citet{77LiPaxx.YO} and \citet{17ZhZhZh.YO}.

Our high-level \ai\ study \citep{19SmSoYu} forms a prerequisite for this work, where a mixture of multireference configuration interaction (MRCI) and coupled cluster methods were used to produce potential energy cures (PECs),  spin--orbit curves (SOCs), electronic angular momentum curves (EAMCs), electric dipole moment curves (DMCs), and transition dipole moment curves (TDMCs) covering the six lowest electronic states of YO. Other theoretical studies of YO include MRCI calculations by \citet{88LaBaxx.YO} and CASPT2 calculations of spectroscopic constants by \citet{17ZhZhZh.YO}.

In this paper,  the \ai\ spectroscopic model of \citet{19SmSoYu} is extended by introducing  non-adiabatic coupling curves for two pairs of states, $A$/$C$ and $B$/$D$, and then refined by fitting to  experimentally derived energies of YO using our coupled nuclear-motion program \Duo\ \citep{Duo}. The energies are constructed  using a combination of the spectroscopic constants and line positions taken from the literature through a procedure based on the MARVEL \citep{MARVEL}  methodology. The new empirical spectroscopic model is used to produce the hot line list \name\ for three major isotopologues of YO, $^{89}$Y$^{16}$O, $^{89}$Y$^{17}$O  and $^{89}$Y$^{18}$O  as part of the ExoMol project \citep{jt528,jt810}.

\section{Experimental information}

Although the spectroscopy of YO has been extensively studied, some key high resolution experimental sources from the 1970-80s only provide spectroscopic constants rather than original transition frequencies; this limits  their usability for high resolution applications. For cases where only spectroscopic constants are available we used an effective Hamiltonian model to compute  the corresponding energy term values. This includes term values for the ground electronic \XS\ state. In the following,  experimental studies of YO  are reviewed.

\textbf{61UhAk}: \citet{61UhAkxx.YO} reported line positions from the \BS--\XS\ (0,0) and \AS--\XS\ (0,0) bands, but   the \BS--\XS\ band is fully covered by more recent and accurate data \citep{80BeGrxx.YO}. The quantum numbers $F_1$ and  $F_2$  of  \BS--\XS\  had to be swapped to agree with  \citet{80BeGrxx.YO}. 
However, due to many conflicting combination differences, only high $J$ transitions ($J>100.5$) were included in our final analysis.


\textbf{77ShNi}: \citet{77ShNixx.YO} performed an analysis of the blue-green \BS--\XS\ and orange  \AS--\XS\ systems  but no rovibronic assignment was reported and their data are not used here.

\textbf{79BeBaLu}: \citet{79BeBaLu.YO} reported an extensive analysis of the $A$--$X$ ($v'=0,1,2,3,4,5$) and $B$--$X$ ($v'=0,1$) systems. Only spectroscopic constants were reported. This work has been superseded by more recent studies and therefore is not used here.

\textbf{80BeGr}: \citet{80BeGrxx.YO} reported line positions from the \BS--\XS\ system, $v'=0,1$ and $v''=0,1,2,3$ and spectroscopic constants for $v'=0,1,2,3$, in emission in a hollow cathode discharge with a partial analysis of the $(3,2)$ and $(4,3)$ bands (only higher $J\ge35.5$ and $J\ge 57.5$, respectively). The data were included in our analysis.


\textbf{83BeGr}: \citet{83BeGrxx.YO} presented a study of the \AS-\XS\ system of YO excited in the discharge of a hollow cathode tube.  Only spectroscopic constants were reported, covering the \XS\ ($v=0\ldots 10$) and \AS\ ($v=0\ldots 9$) vibronic states ($v= 6$ and $v=7$ with a limited analysis). We used these constants and the effective Hamiltonian of \citet{77BaCeDI.CN} to  generate term values for the states $v=0\ldots 5$, 8 and 9 (\AS). It should be noted that the spectroscopic program \pgopher\ \cite{PGOPHER} could not be used as its $^2\Pi$ model is found to be incompatible with that used by \citet{83BeGrxx.YO} despite constants sharing the same names. A simple Python code based on the  effective Hamiltonian expressions of \citet{77BaCeDI.CN} is provided as part of our supplementary material.  Band heads of $v'=7-15$ ($\Omega= 0.5$) and $v'=6,8,9$ ($\Omega=1.5$) were reported, but not used directly in the fit here. It is known that effective Hamiltonian expansions can diverge at high $J$,  we therefore limited the corresponding energies to about $J \leq 100.5$.

The coverage of the spectroscopic constants of \XS\ is up to $v=10$, while the available line positions are only up to $v=3$, which is why we opted to use the spectroscopic constants by \citet{83BeGrxx.YO} to generate the  \XS\ term values. YO has a relatively rigid structure in its ground electronic state potential with the vibronic energies well separated from each other and other electronic state.

\textbf{92SiJaHa}: \citet{92SiJaHa.YO} reported line positions from the \ApS--\XS\ system (0,0) in their laser-induced fluorescence spectral study with a pulsed dye laser. It was included in the analysis here.

\textbf{93HoTo}: \citet{93HoToxx.YO} reported a microwave spectrum of \XS\ for $v=0,1,2,3$. It was included in the analysis here.

\textbf{05LeMaCh}: \citet{05LeMaCh.YO} reported a cavity ring-down absorption spectrum of the \BS--\XS\  (2,0) and (2,1) system. We included their line positions in the analysis here.

\textbf{15CoHuYe}: \citet{15CoHuYe.YO} reported three THz lines from the \AS--\XS\ $(0,0)$ system with low uncertainties recorded for laser cooling application. These were included in our analysis.

\textbf{17ZhZhZh}: \citet{17ZhZhZh.YO} reported line positions from the \DS--\XS\ system (0,0) and (1,0) which were used in the analysis here.

\textbf{23MuNa}: \citet{23MuNa.YO}  reported a high-resolution analysis of the highly excited \AS--\XS\ ($v'=11,12,13$) and \BS--\XS\  ($v' = 5,6$) systems. For the $A$--$X$ band, only the $\Omega=0.5$ branch was provided. Their line positions were included in our analysis. There is a crossing between \AS, $v=11$ and \DS, $v=2$ around $J=34.5$, see Fig.~\ref{f:A:D:cross}.

The only experimental information on the transition probabilities available for  YO includes the permanent dipole moments in  the \XS\ and \AS\ states  measured  by \citet{90StShxx.YO,90SuLoFr.YO,03StVixx.YO} and the lifetimes of some lower lying vibrational states  measured by \citet{77LiPaxx.YO} (\AS) and \citet{17ZhZhZh.YO} (\BS\ and \DS).



\section{Description of the pseudo-MARVEL procedures}

MARVEL (Measured Active Rotational Vibrational Energy Levels) is a spectroscopic network algorithm \citep{MARVEL}, now routinely used for constructing  ExoMol line lists for high-resolution applications \citep{jt810}.
We did not have sufficient original experimental line positions for a proper MARVEL analysis of the YO spectroscopic data, which were mostly only available represented by spectroscopic constants.   Furthermore, there are no studies of the infrared spectrum of YO which meant that the (lower) ground energies could  be only reconstructed from lower quality UV transitions, which limits the quality of the MARVEL energies.


Instead, a `pseudo-MARVEL' procedure was applied as follows (see also \citet{22YuNoAz}). The  experimental frequencies $\tilde{\nu}_{ij}$, where available, were utilised to generate rovibronic energies of YO as upper states energies using
\begin{equation}
\label{e:E'=nu+E''}
\tilde{E}_{j(i)}^{\rm (upp)}= \tilde{E}^{\rm (low)}_{i} + \tilde{\nu}_{ij},
\end{equation}
which were then averaged over all transitions connecting the same upper state $j$. All experimental transitions originate from or end up at the \XS\  state. We used the spectroscopic constants from \citet{83BeGrxx.YO} ($v=0,1,\ldots,10$) to generate the  \XS\ state energies $\tilde{E}^{\rm (low)}_{i}$ in conjunction with the program \pgopher. The \AS\ state energies were generated using  the effective Hamiltonian \citep{77BaCeDI.CN} except for $v=0$, $J> 100.5$, which were obtained using the pseudo-MARVEL procedure. This pseudo-MARVEL analysis yielded 5089 empirically determined energy levels which we used in the fit. The final experimentally determined energy set covers the following vibronic bands $X$: $v= 0-10$; $A'$: $v=0$, $J_{\rm max} = 14.5$;  $A$: $v=$ 0,1,2,3,4,5, 8,9, 11,12,13; $B$: $v=0,1,2,3,4,5,6$; $D$: $v = 0,1$. The vibrational and rotational coverage is illustrated in Fig.~\ref{f:obs}.  The experimental transition frequencies collected as part of this work are provided in  the Supporting Information to this paper in the MARVEL format together with the pseudo-MARVEL energies used in the fit.

\begin{figure*}
    \centering
    \includegraphics[width=0.44\textwidth]{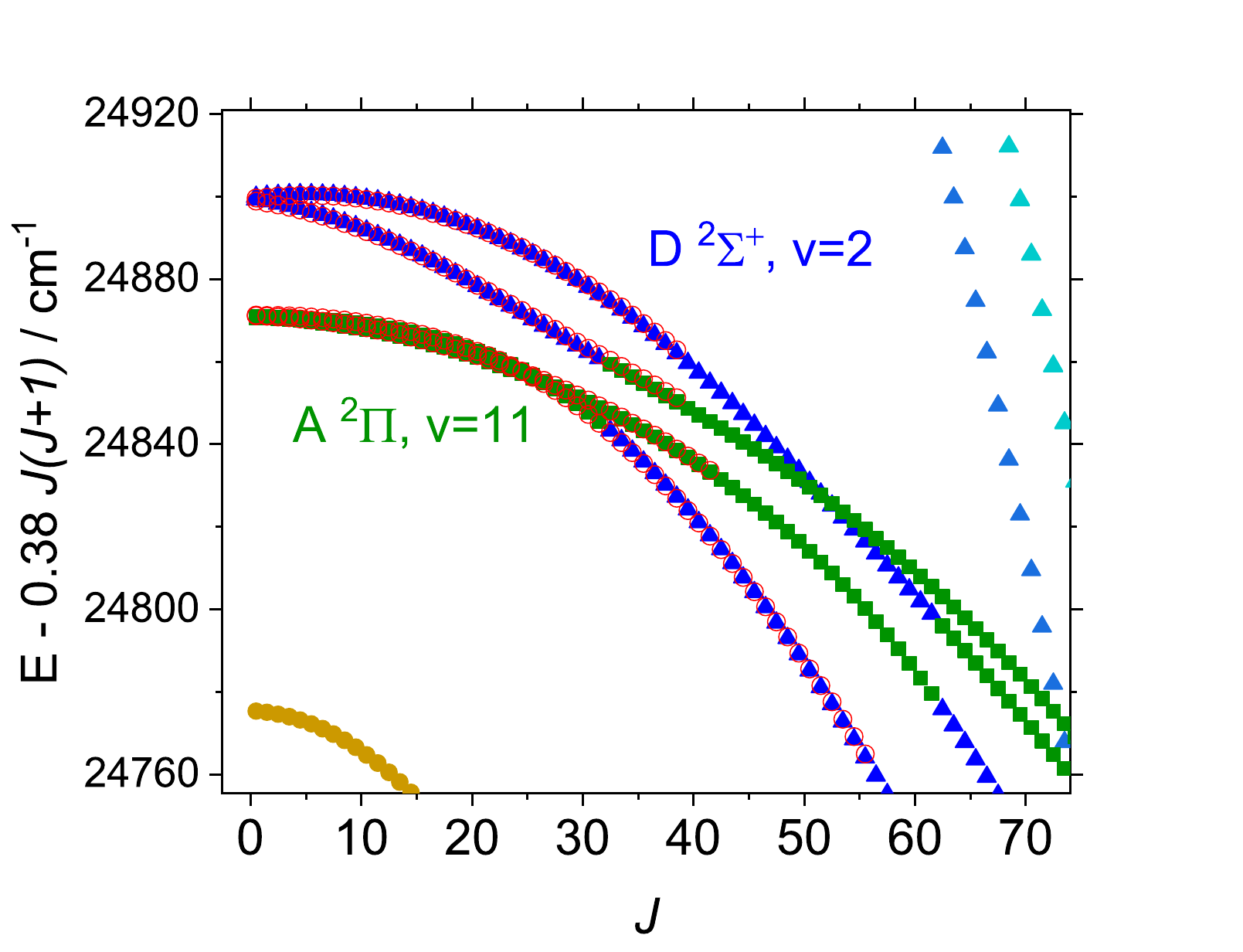}
    \includegraphics[width=0.44\textwidth]{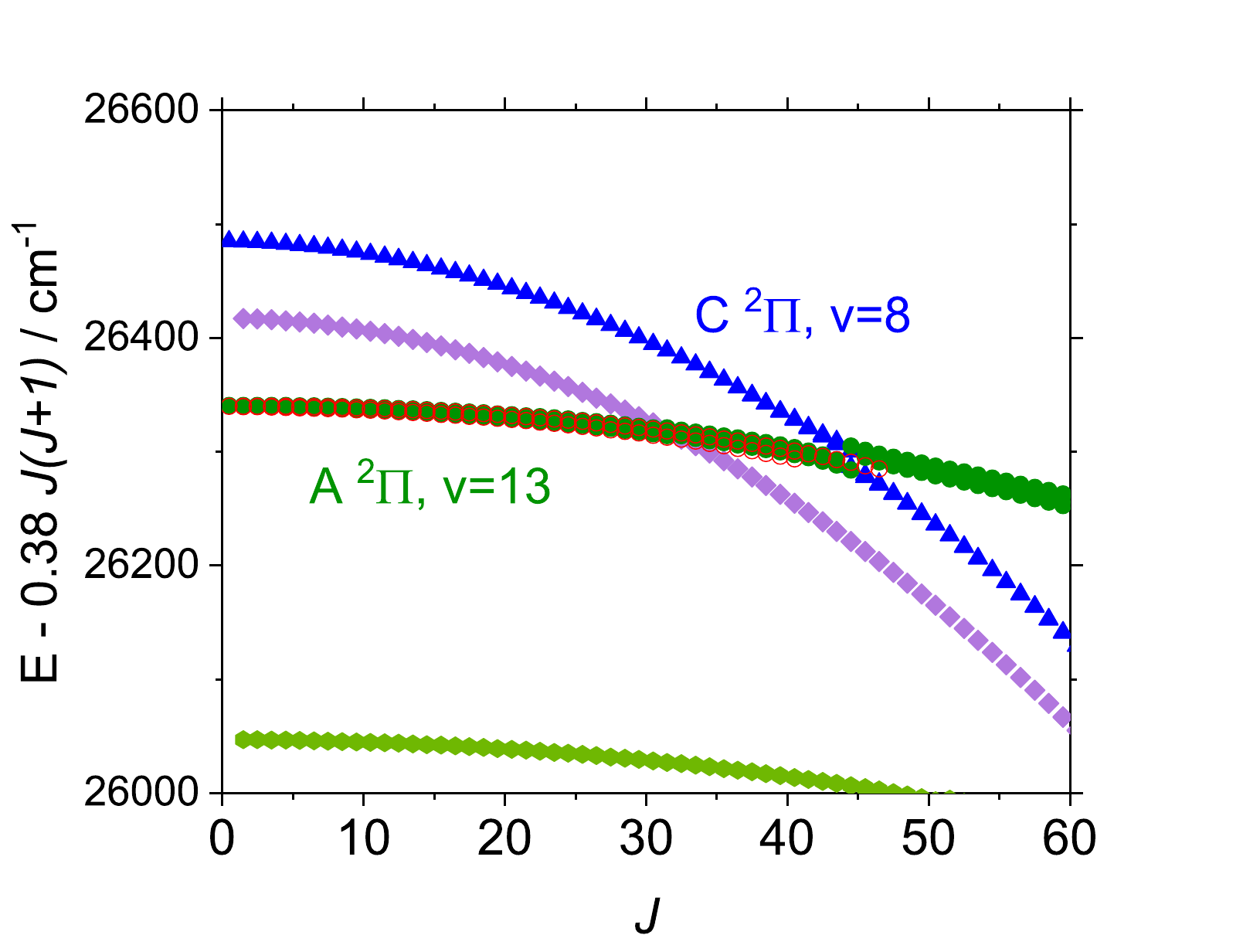}
 \caption{Illustration of the crossings: \DS, $v=2$ and  \AS, $v=11$ bands (left) and  \AS, $v=13$ and \CS, $v=8$ (right), with experimental values shown as empty circle. }
    \label{f:A:D:cross}
\end{figure*}

\begin{figure}
    \centering
    \includegraphics[width=0.44\textwidth]{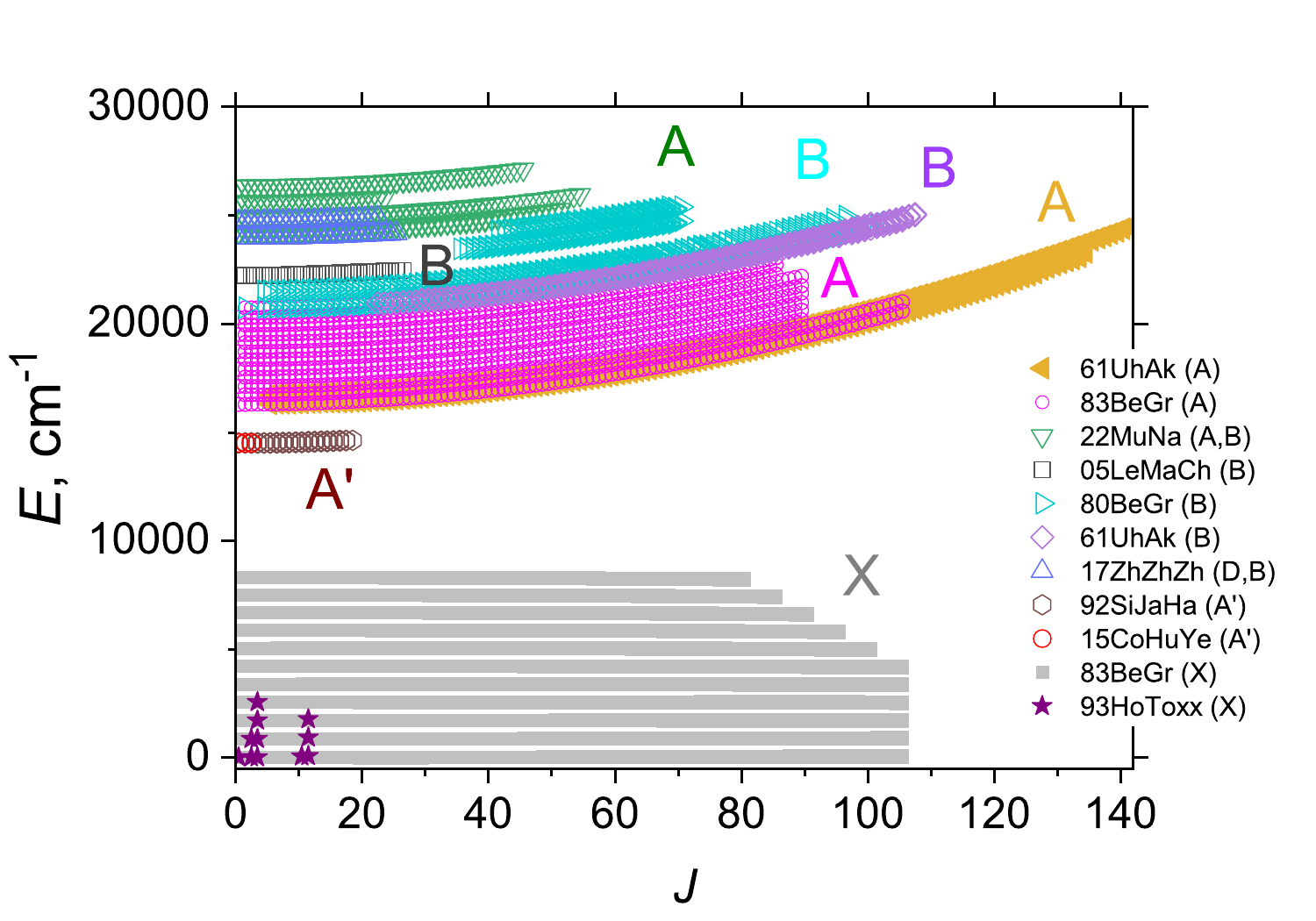}
\caption{Experimentally derived energy term values of YO used in the refinement of the \ai\ spectroscopic model.}
    \label{f:obs}
\end{figure}

It should be noted that  the effective Hamiltonians used  do not provide any information on  direct perturbations between vibronic bands caused by their crossing or any other inter-band interactions. For example, the \AS\ $v=5$ and \BS\ $v=0$ states cross at around $J=27.5$. We excluded   energy values in the  vicinity of such crossings from the fit.  The only  crossing  represented by the real data  is  between the \DS\ $v=2$ and \AS\ $v=11$ bands \citep{23MuNa.YO}, see Fig.~\ref{f:A:D:cross}.

\section{\textit{Ab initio} calculations}

\begin{figure*}
    \centering
    \includegraphics[width=0.44\textwidth]{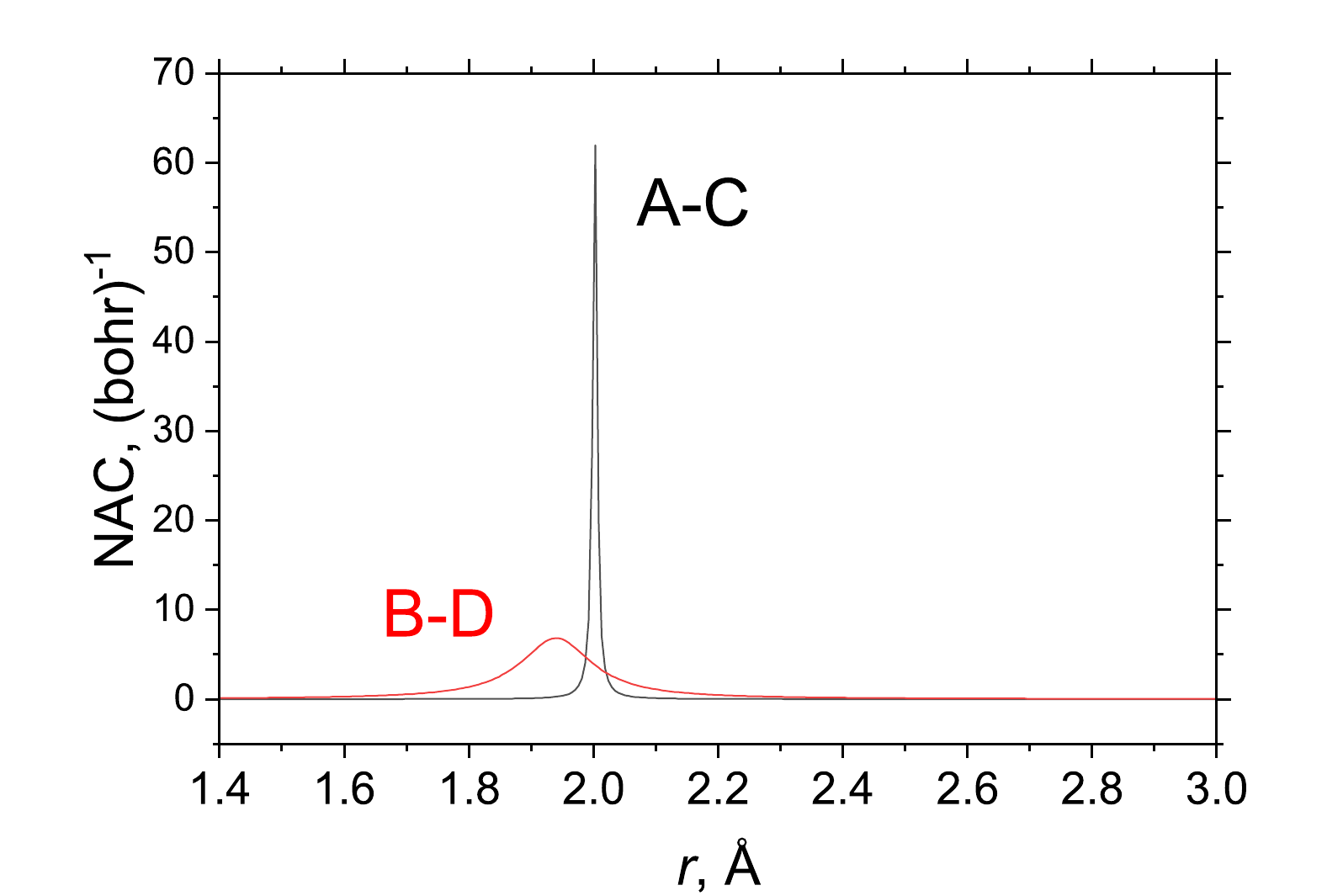}
    \includegraphics[width=0.44\textwidth]{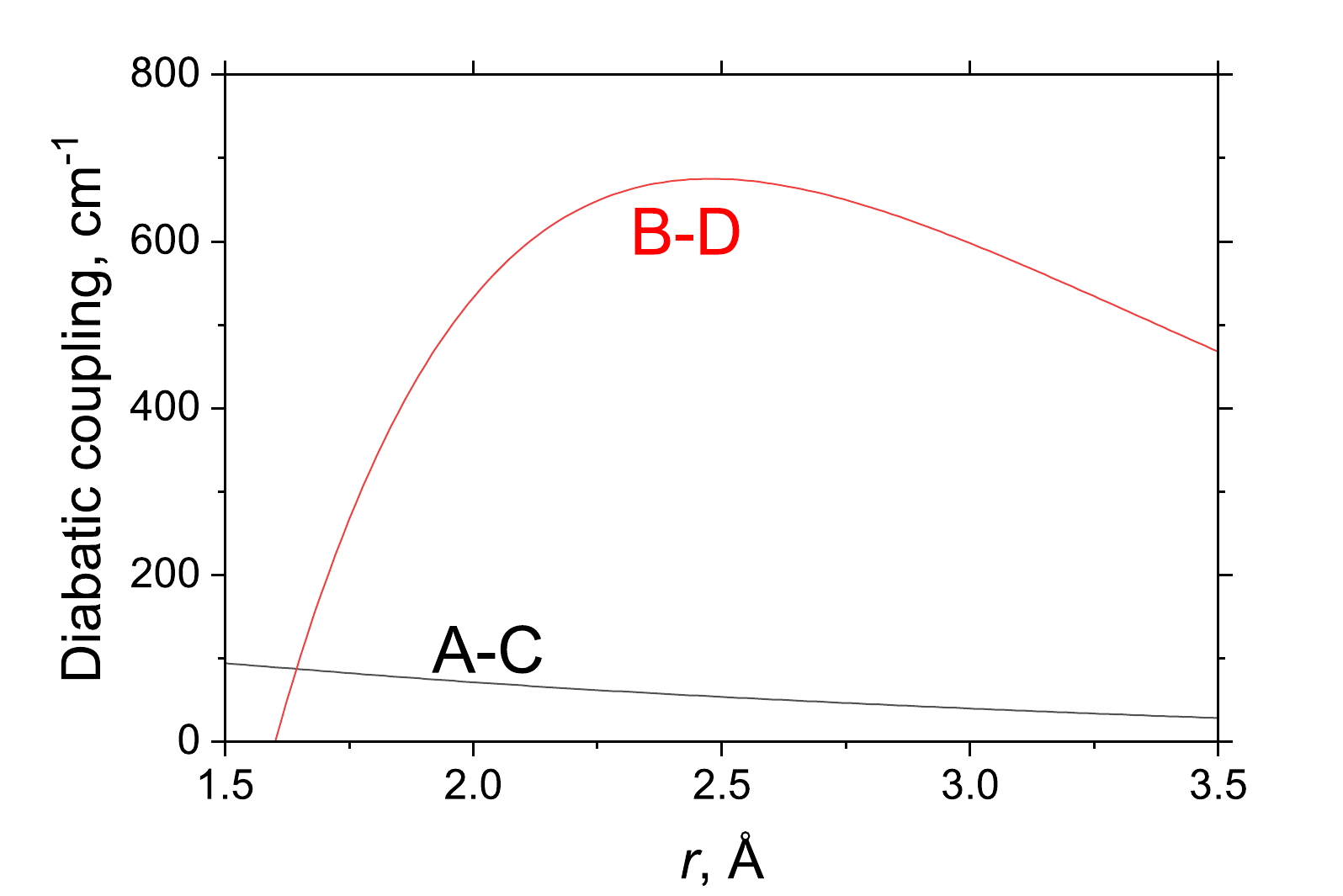}
\caption{CASSCF NACs and empirical diabatic couplings $D(r)$ of YO, A--C and B--D.}
\label{f:NAC}
\end{figure*}

Non-adiabatic couplings (NACs) or the first-order derivative couplings between the state pairs $\XSM-\BSM$, $\XSM-\DSM$, $\BSM-\DSM$\ and $\ASM-\CSM$\  were derived by three-point central differences for CASSCF wavefunctions using the DDR
procedure as implemented in MOLPRO \citep{MOLPRO2020}.
The state-averaged CASSCF method was employed with density matrix averaging over six low-lying doublet states (three $\Sigma^+$, two $\Pi$, and one $\Delta$) with equal weights for each of the roots. The active space included 7 electrons distributed in 13 orbitals (6$a_1$, 3$b_1$,  3$b_2$, 1$a_2$) that had predominantly oxygen 2p and yttrium 4d, 5s, 5p, and 6s character; all lower energy orbitals were constrained to be doubly occupied. Augmented triple-zeta quality basis sets aug-cc-pwCVTZ \citep{02PeDuxx.ai} on O and pseudopotential-based aug-cc-pwCVTZ-PP \citep{07PeFiDo.ai} on Y were used in these calculations. The resulting NACs are illustrated in Fig.~\ref{f:NAC}.

\section{Spectroscopic model}

Our starting point is the \ai\ spectroscopic model of YO developed by \citet{19SmSoYu}, which includes PECs, SOCs, TDMCs, EAMCs for six lowest doublet  states of YO in the adiabatic representation. YO exhibits non-adiabatic effects via the couplings of the two pairs of states: \AS\ with \CS\ and \BS\ with  \DS. Apart from the avoided crossing in these PECs, other  adiabatic curves (SOCs, EAMCs, (T)DMCs) also have   strongly distorted shapes exhibiting step-like behaviour, which makes the adiabatic representation far from ideal for refinement. This is not only because these curves are difficult to represent analytically as parameterised functions of the bond length (required for the fit), but also because the shapes of the curves around any avoided crossing are very sensitive to the precise position of these crossings, which are also difficult to control in the adiabatic representation.
Due to these effects, SOCs, EAMCs and (T)DMCs  between the \BS, \DS\ and the \AS, \CS\ states also exhibit discontinuous behaviour in the region of the avoided crossing, which can only be correctly treated in combination with the NAC curves  as well as their second-order derivative couplings (which we did not compute). Vibronic intensities, for instance,  are very sensitive to the description of the steep structures in the adiabatic DMCs. Because of inaccuracies in the \ai\ calculations, the adiabatic DMCs will be prone to large errors in their shape since the strong, steep gradient variations around the avoided crossing are sensitive to both crossing position and morphology, and so will negatively affect the corresponding spectral properties. These sharp behaviours are difficult to model, so the diabatic representation is a natural choice since DMCs and other couplings will become smooth, and less sensitive to inaccuracies in \ai\ calculations.

We therefore  decided to work in the diabatic representation taking advantage of the recent developments in \duo\ \citep{22BrYuKi,23BrDrTe}. To this end, a diabatic spectroscopic model of YO was generated by diabatising the \ai\ adiabatic PECs, SOCs, EAMCs and (T)DMCs of YO \citep{19SmSoYu} as outlined in the following.

Unfortunately, the \ai\ adiabatic curves reported in \citet{19SmSoYu} were not suitable for a direct diabatisation using the corresponding NACs due to  incomplete \ai\ curves and inconsistent levels of theory used for different properties. Effectively, only the  PECs of the six electronic states of YO computed using the complete basis set limit (CBS) extrapolation from awCVQZ and awCV5Z in conjunction with the CCSD(T) method  were suitable for  accurate descriptions of the corresponding crossings in the diabatic  representations. All other property curves (SOCs, EAMCs, (T)DMCs)  were computed with MRCI or even CASSCF and did not provide  adequate coverage, especially at longer bond lengths ($r>1.9$~\AA)  beyond the avoided crossing points (see Figs.~9 and 10 in \citet{19SmSoYu}).

In order to overcome this problem, in line with the property-based diabatisation (see, e.g. \citet{22ShVaZo}),  we constructed diabatic curves under the assumption  that in the diabatic representations all the curves become smooth, without  characteristic step-like shapes  and manually constructed diabatic SOCs, EMACs, TDMCs. The existing points  were inter- and extrapolated to best represent smooth diabatic curves.  Admittedly, there is some arbitrariness in this approach which is subsequently resolved, at least partially, by empirically refining the initial curves.  The various curves representing  our diabatic spectroscopic model of YO are illustrated in Figs.~\ref{f:SOCs}--\ref{f:Dipoles}.   

\subsection{Diabatisation}
\label{s:diab}

To represent the diabatic potential energy curves of the \XS, \AS, \ApS,  \BS, \CS\ and \DS\ states analytically, we used the extended Morse oscillator (EMO)   \citep{EMO} function as well as the  Extended Hulburt-Hirschfelder (EHH) function  \citep{41HuHi} as implemented in \Duo. An EMO function is given by
\begin{equation}\label{e:EMO}
V(r)=V_{\rm e}\;\;+\;\;(A_{\rm e} - V_{\rm
e})\left[1\;\;-\;\;\exp\left(-\sum_{k=0}^{N} B_{k}\xi_p^{k}(r-r_{\rm e})
\right)\right]^2,
\end{equation}
where $A_{\rm e} $ is a dissociation asymptote,  $A_{\rm e} - V_{\rm e}$ is the dissociation energy, $r_{\rm e}$ is an equilibrium distance of the PEC, and $\xi_p$ is the \v{S}urkus variable given by:
\begin{equation}
\label{e:surkus:2}
\xi_p= \frac{r^{p}-r^{p}_{\rm e}}{r^{p}+r^{p}_{\rm e }}.
\end{equation}
The EMO form is our  typical choice for representing PECs and was used here for the  \XS, \AS, \BS\ and \DS\ states. For the \ApS\ and \CS\ states, which do not have much experimental information for refinement, we employ the  EHH function. This form was suggested to be more suitable for the description of the dissociation region \citep{06CaClLi.PN}. Here we use the EHH form from  \citet{23UsSeYu} as given by
\begin{equation}
    V_{\textrm{EHH}}(r)=D_\textrm{e}\left[\left(1-e^{-q}\right)^2 + cq^3\left(1+\sum_{i=1}^3 b_iq^i\right)e^{-2q}\right], \label{EHH}
\end{equation}
where $q=\alpha \left(r-r_\textrm{e}\right)$.

The corresponding  parameters defining PECs were first obtained by  fitting to the \ai\ CCSD(T)/CBS potential energies  and then empirically refined by fitting to empirical energies of YO (where available) as described below; these parameters are  given in the supplementary material in the form of a Duo input file. The asymptotic energies $A_{\rm e}$ for all states but \BS\ were fixed to the value  59220~\cm,  or 7.34~eV, which corresponds to $D_0$ = 7.290(87) eV determined  by \citet{74AcRa}, based on their mass spectrometric measurements. For the \BS\ state, $A_{\rm e}$ was fixed to a higher value  of 75000 \cm\ in order to achieve a physically sensible shape of the PEC. Otherwise, the \BS\ curve tended to cross the \DS\ curve also at $r \sim 4$~\AA.


In principle, the property-based diabatisation does not require  the usage of the NAC curves. However, in order to assist our diabatisations of the YO  \ai\ curves, we used the \ai\ CASSCF NACs of  $A$--$C$ and $B$-$D$  shown in Fig.~\ref{f:NAC} as a guide. These curves were fitted using the following Lorentzian functions:
\begin{equation}
\phi_{ij}(r) =  \frac{1}{2}\frac{\gamma}{\gamma^2+(r-r_{\rm c})^2},
\label{e:lorentzian}
\end{equation}
where $\gamma$ is the corresponding half-width-at-half-maximum (HWHM), while $r_{\rm c}$ is its centre, corresponding to the crossing point of diabatic curves.

The diabatic and adiabatic representations are connected via a unitary $2\times 2$ transformation given by
\begin{equation}
{\bf U}(\beta(r))= \begin{bmatrix} \cos(\beta(r)) & -\sin(\beta(r)) \\ \sin(\beta(r)) & \cos(\beta(r)) \end{bmatrix},
\label{e:U(beta)}
\end{equation}
where the $r$-dependent mixing angle $\beta(r)$ is obtained via the integral
\begin{equation}
\label{e:beta(r)}
\beta(r)= \int^{r}_{-\infty} \phi_{12}(r') dr' .
\end{equation}
For the Lorentzian-type NAC in Eq.~\eqref{e:lorentzian}, the angle $\beta$ is given by
\begin{equation}
\label{e:beta:Lor}
\beta=\frac{1}{2}\arctan\left(\frac{r-r_{\rm c}}{\gamma}\right)+\frac{\pi}{4}.
\end{equation}

The  diabatic representation is defined by two PECs $V_1(r)$  and $V_2(r)$ coupled with a diabatic term $D(r)$ as a  $2\times 2$ diabatic matrix
\begin{equation}\label{e:V1W/WV2}
\bf{A} = \left(
\begin{array}{cc}
  V_1(r) & D(r) \\
  D(r) & V_2(r)
\end{array}
\right).
\end{equation}
The two eigenvalues  of the matrix $\bf{A}$ provide the adiabatic PECs in the form of  solution of a quadratic equation as given by
\begin{eqnarray}
  V_{\rm low}(r) &=& \frac{V_1(r)+V_2(r)}{2}-\frac{\sqrt{[V_1(r)-V_2(r)]^2+4 \, D^2(r)}}{2}, \\
  V_{\rm upp}(r) &=& \frac{V_1(r)+V_2(r)}{2}+\frac{\sqrt{[V_1(r)-V_2(r)]^2+4 \, D^2(r)}}{2},
\end{eqnarray}
where $V_{\rm low}(r)$ and $V_{\rm upp}(r)$ are the two adiabatic PECs.

Assuming  the diabatic PECs $V_1(r)$ and $V_2(r)$ as well as  NAC $\phi_{12}(r)$ are known, the diabatic coupling function $D(r)$ can be re-constructed using the condition that the non-diagonal coupling should vanish upon the unitary transformation $U(r)$ in Eq.~\eqref{e:U(beta)} such that the adiabatic potential matrix is diagonal and is then given by:
\begin{equation}\label{e:D12}
D(r) = \frac{1}{2}\tan(2\beta(r)) \left(V_2(r)-V_1(r)\right).
\end{equation}
Assuming also the EMO functions for the PECs $V_1(r)$ and $V_2(r)$ as   in Eq.~\eqref{e:EMO} and  the `Lorentzian'-type angle $\beta(r)$ in Eq.~\eqref{e:beta:Lor}, the diabatic coupling curves for YO have an asymmetric Gaussian-like shape, see the right panel of Fig.~\ref{f:NAC}; this is not always the case as the $V_2-V_1$ term in Eq.~(\ref{e:D12}) heavily influences the morphology of $D(r)$.

For the \BS--\DS\ pair, where the experimental data is better represented, in order to introduce some flexibility into the fit, we decided to model the diabatic coupling   by  directly representing it using an inverted EMO function from Eq.~\eqref{e:EMO}. This gives the asymmetric Gaussian-like shape, with the  asymptote $A_{\rm e}$  set to zero and $V_{\rm e}$ representing the maximum of the diabatic coupling $D(r)$. The $A$--$C$ diabatic coupling was modelled using Eq.~\eqref{e:D12} with the two parameter `Lorentzian'-type angle $\beta(r)$ from Eq.~\eqref{e:beta:Lor}.

\subsection{Other coupling curves}

\begin{figure*}
    \centering
    \includegraphics[width=0.46\textwidth]{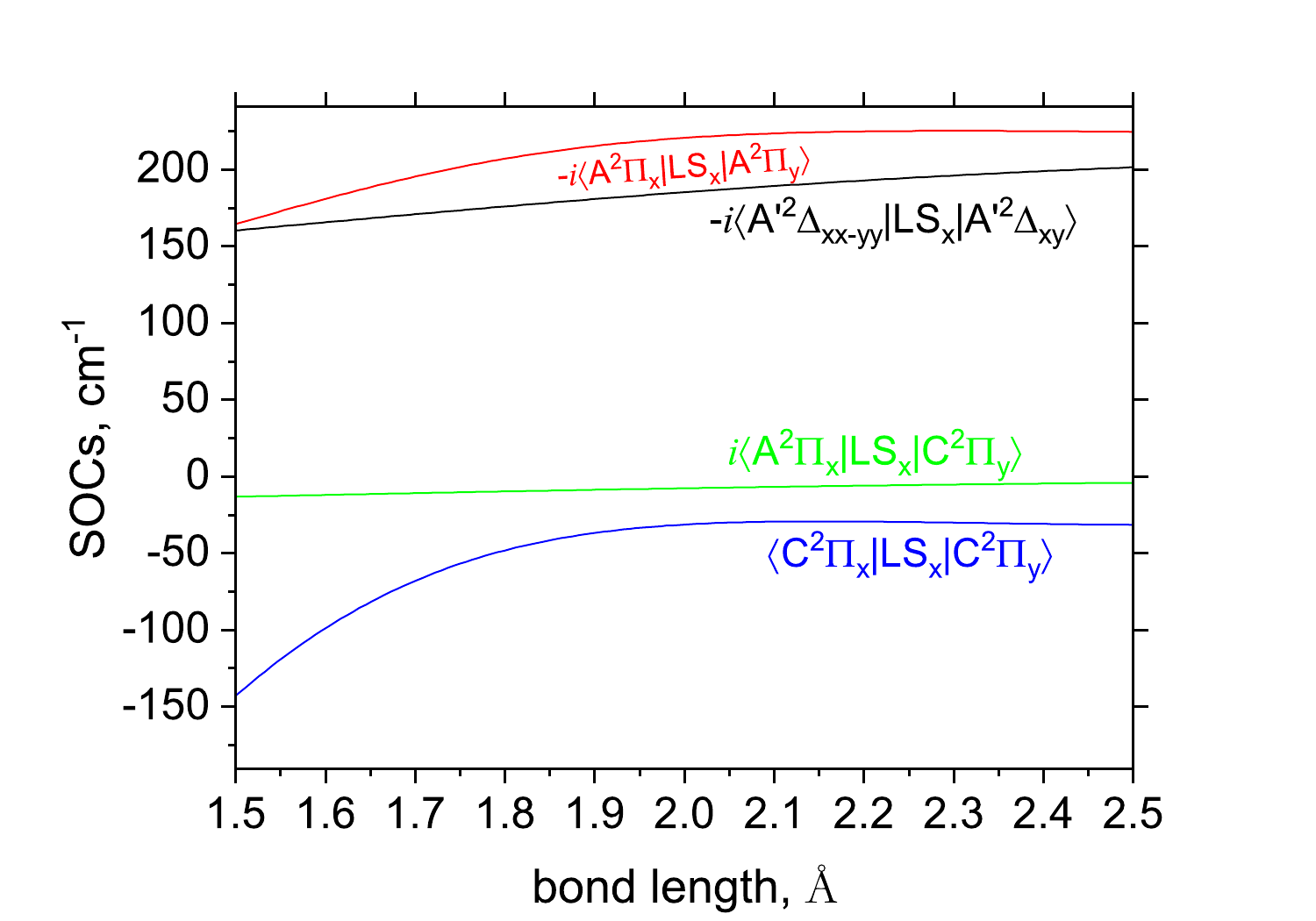}
    \includegraphics[width=0.46\textwidth]{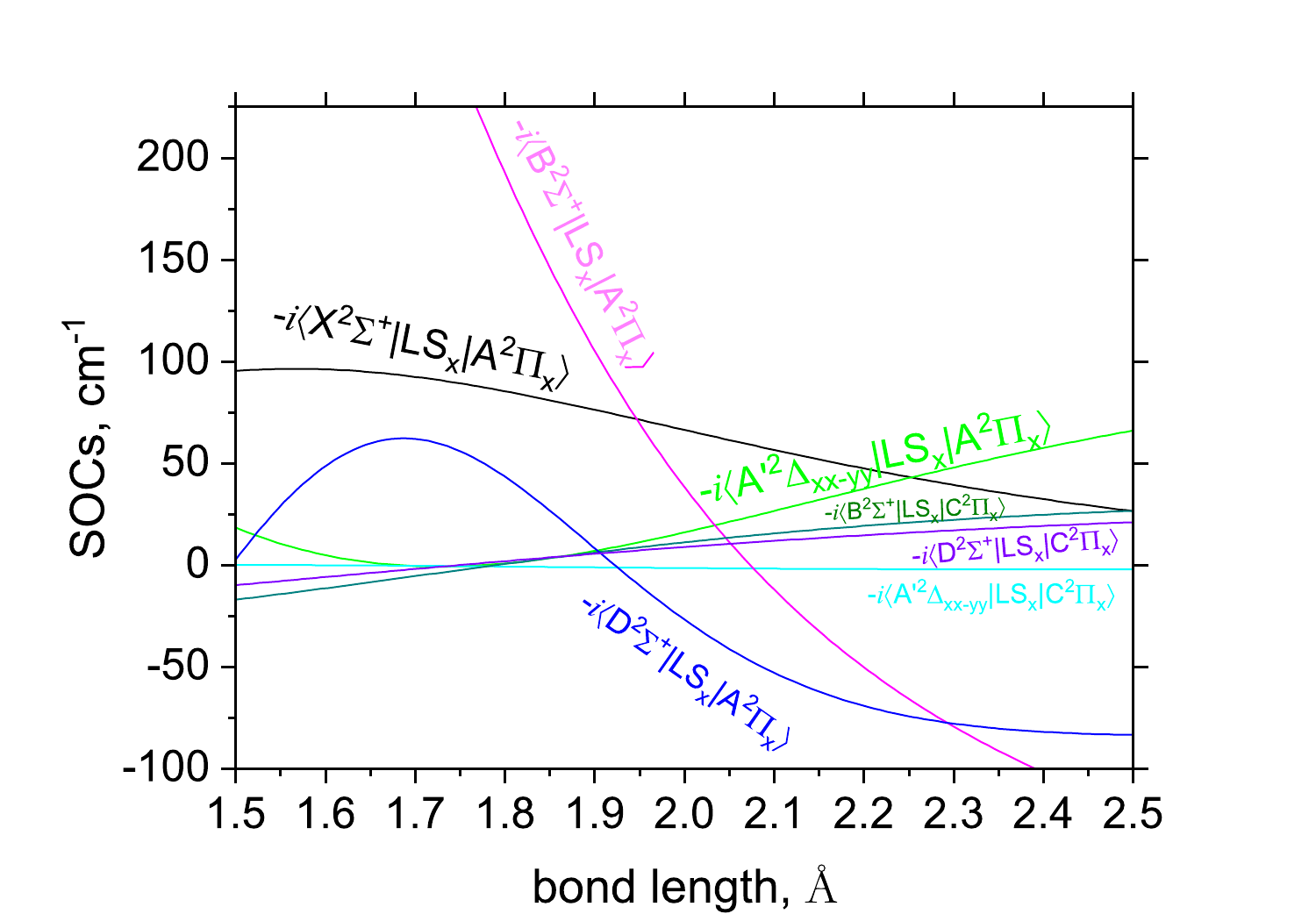}
\caption{Refined SOCs of YO in the diabatic representation: diagonal and non-diagonal.}
    \label{f:SOCs}
\end{figure*}

\begin{figure*}
    \centering
    \includegraphics[width=0.46\textwidth]{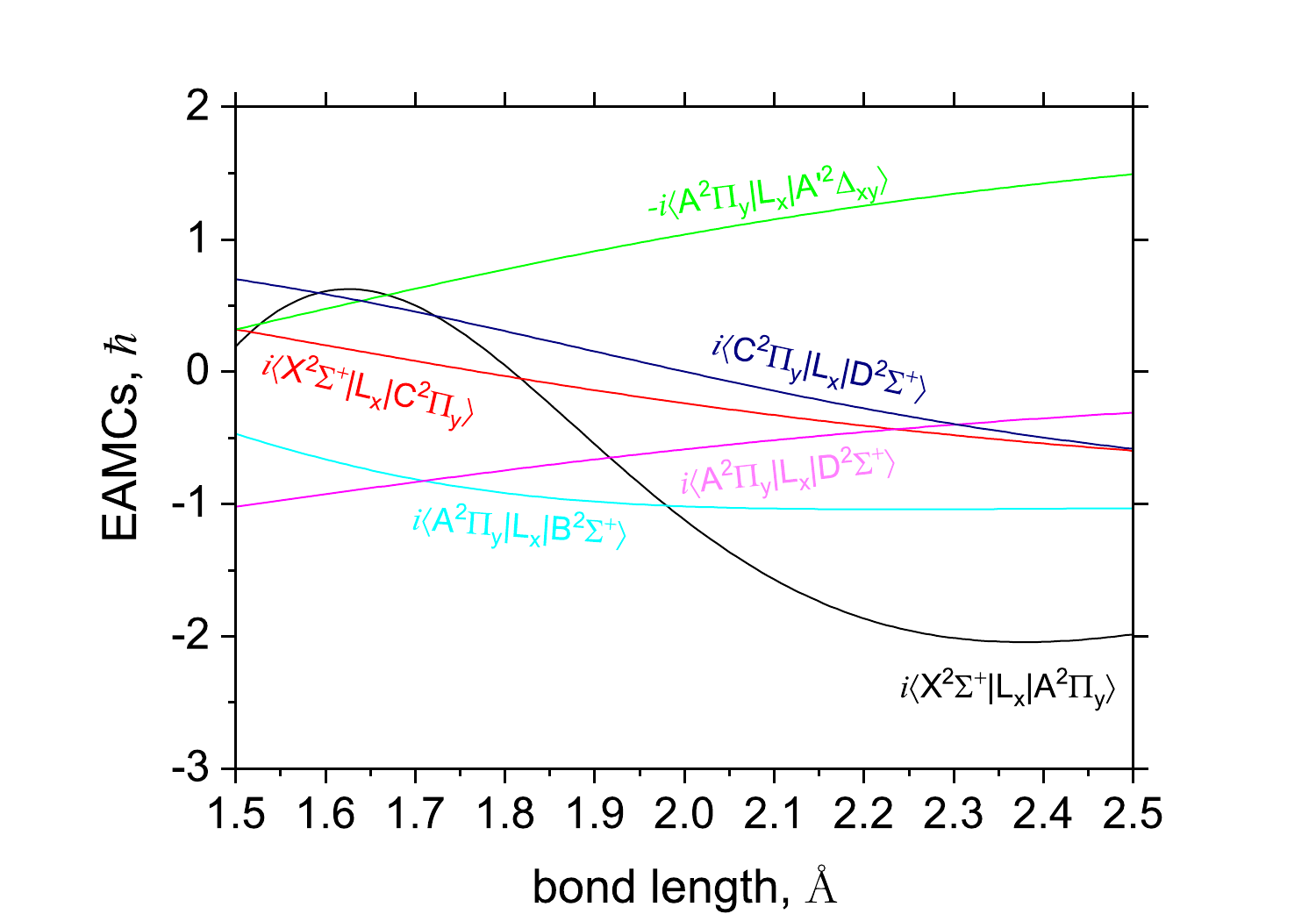}
    \includegraphics[width=0.46\textwidth]{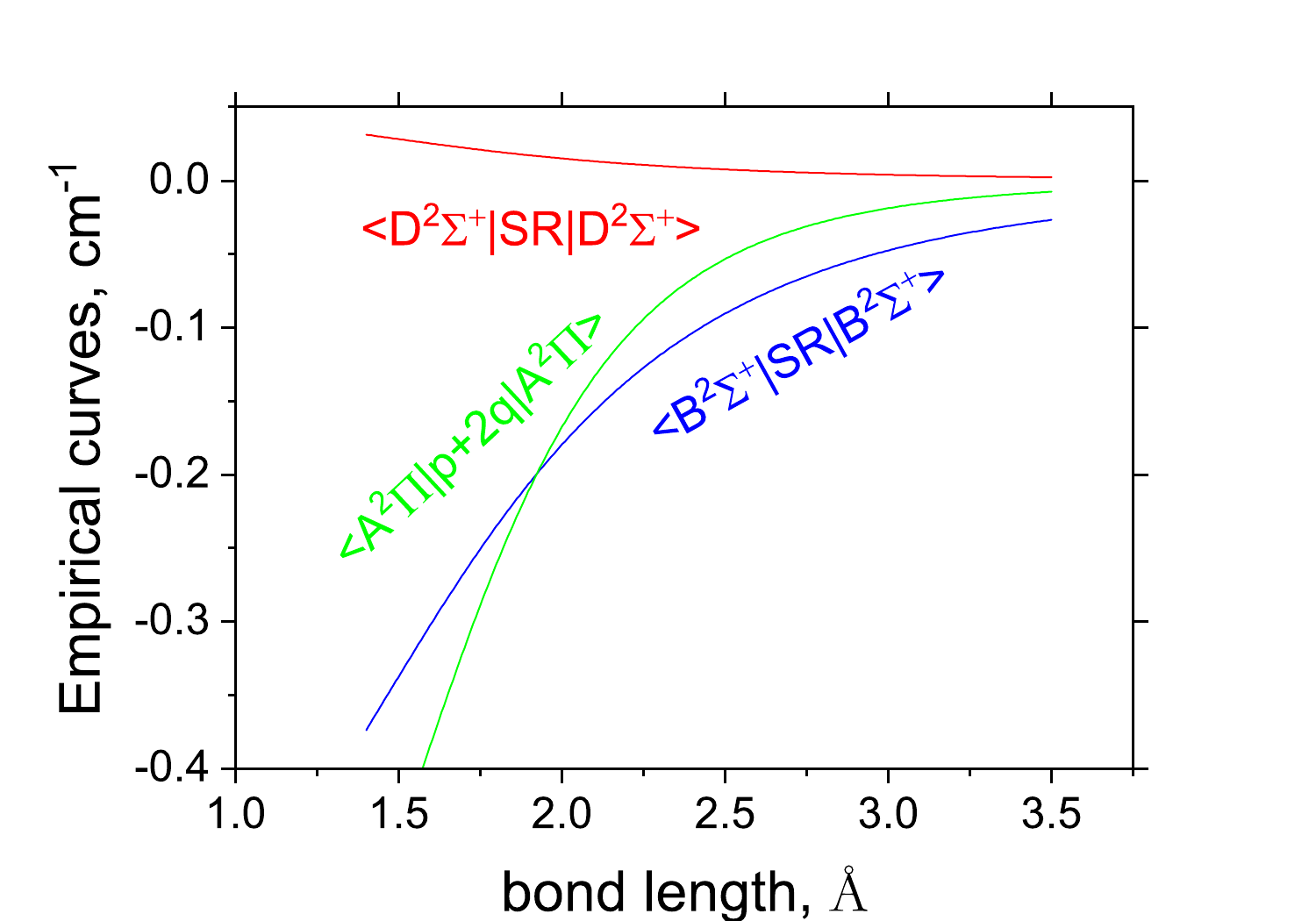}
\caption{Refined  EAMCs of YO in the diabatic representation, empirical spin-rotation corrections and $\Lambda$-doubling curves.}
    \label{f:EMAC}
\end{figure*}

For the SOC and EAMC curves of YO we used the  expansion:
\begin{equation}
\label{e:bob}
F(r)=\sum^{N}_{k=0}B_{k}\, z^{k} (1-\xi_p) + \xi_p\, B_{\infty},
\end{equation}
where $z$ is either taken as the \v{S}urkus variable $z=\xi_p$  or a
damped-coordinate given by:
\begin{equation}\label{e:damp}
z = (r-r_{\rm ref})\, e^{-\beta_2 (r-r_{\rm ref})^2-\beta_4 (r - r_{\rm ref})^4},
\end{equation}
see also \citet{jt703} and \citet{jt711}.  Here $r_{\rm ref}$ is a
reference position equal to $r_{\rm e}$ by default and $\beta_2$ and
$\beta_4$ are damping factors.
For the \XS\ state, a BOB (Born-Oppenheimer Breakdown) correction curve modelled using Eq.~\eqref{e:bob} was used. These parameterised representations were then used to refine the \ai\ curves by fitting them to the experimentally derived rovibronic energies of YO. The final coupling curves are shown in Figs.~\ref{f:NAC} (right display), \ref{f:SOCs} and \ref{f:EMAC}.

We also included spin-rotation and $\Lambda$-doubling $p(r)+2o(r)$ \citep{79BrMexx.methods} curves as empirical objects for some of the electronic states modelled using Eq.~\eqref{e:bob}, see Fig.~\ref{f:EMAC}.

\subsection{Dipoles}
\label{s:dipoles}

\begin{figure}
    \centering
    \includegraphics[width=\linewidth]{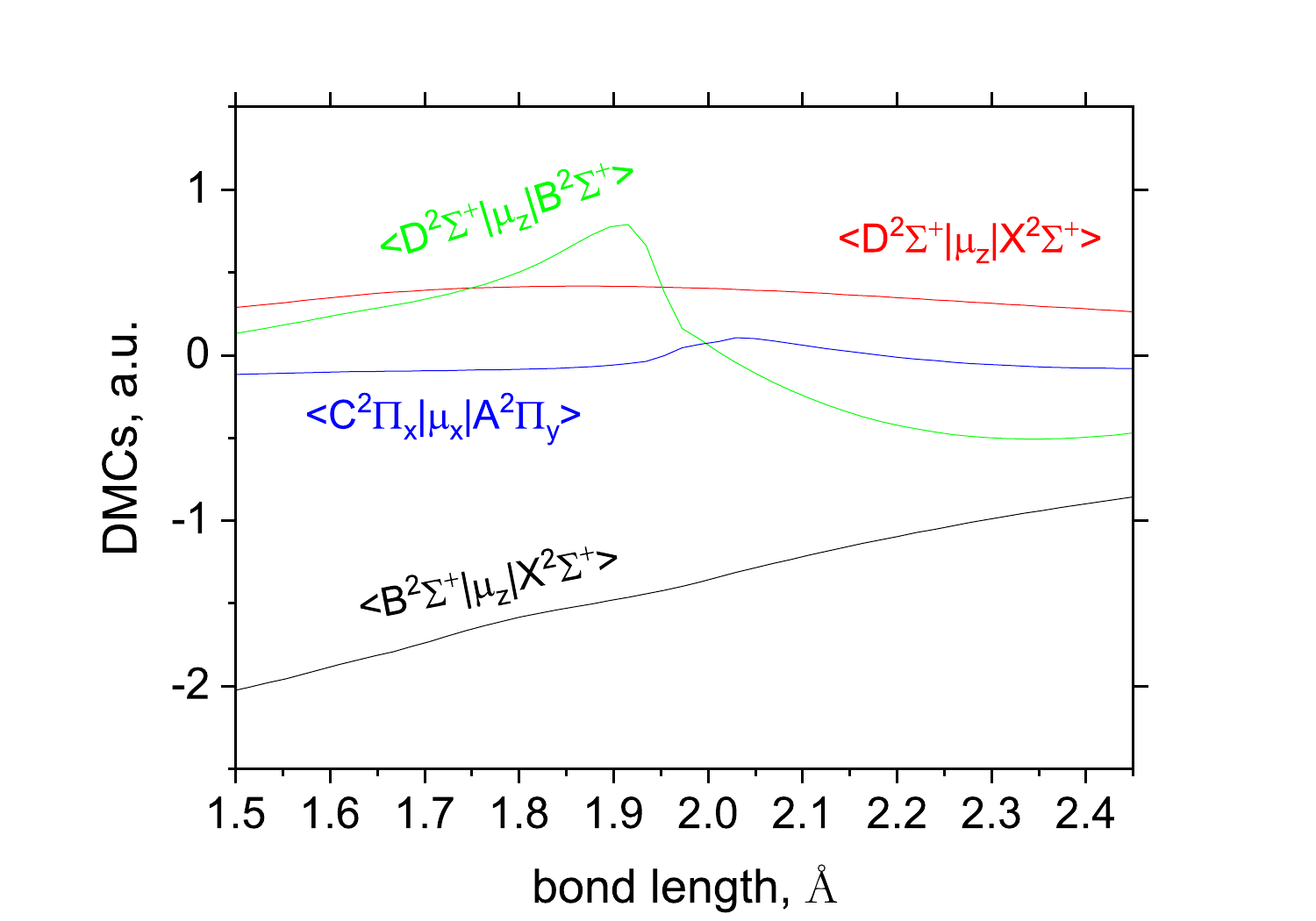}
    \includegraphics[width=\linewidth]{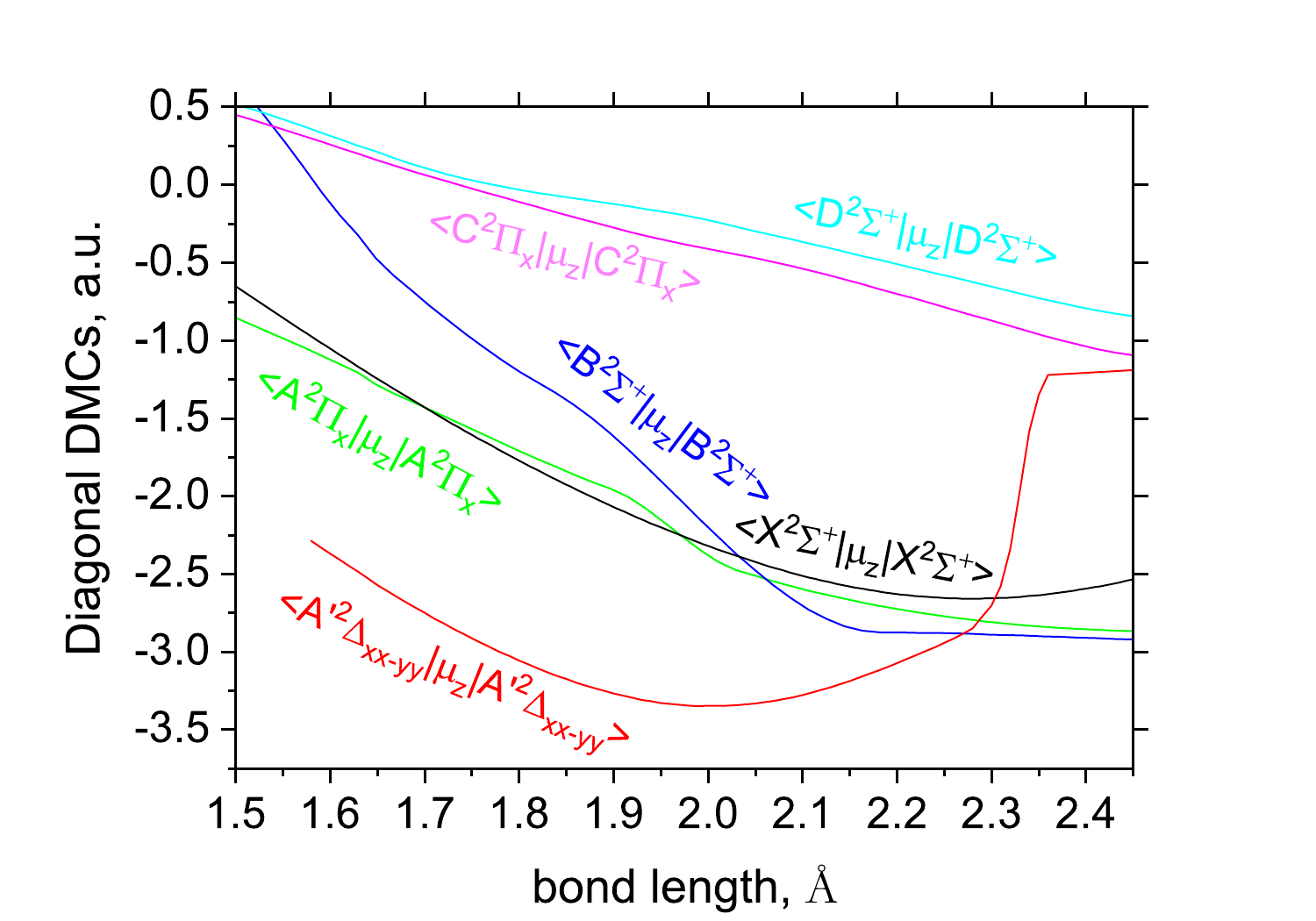}
\includegraphics[width=\linewidth]{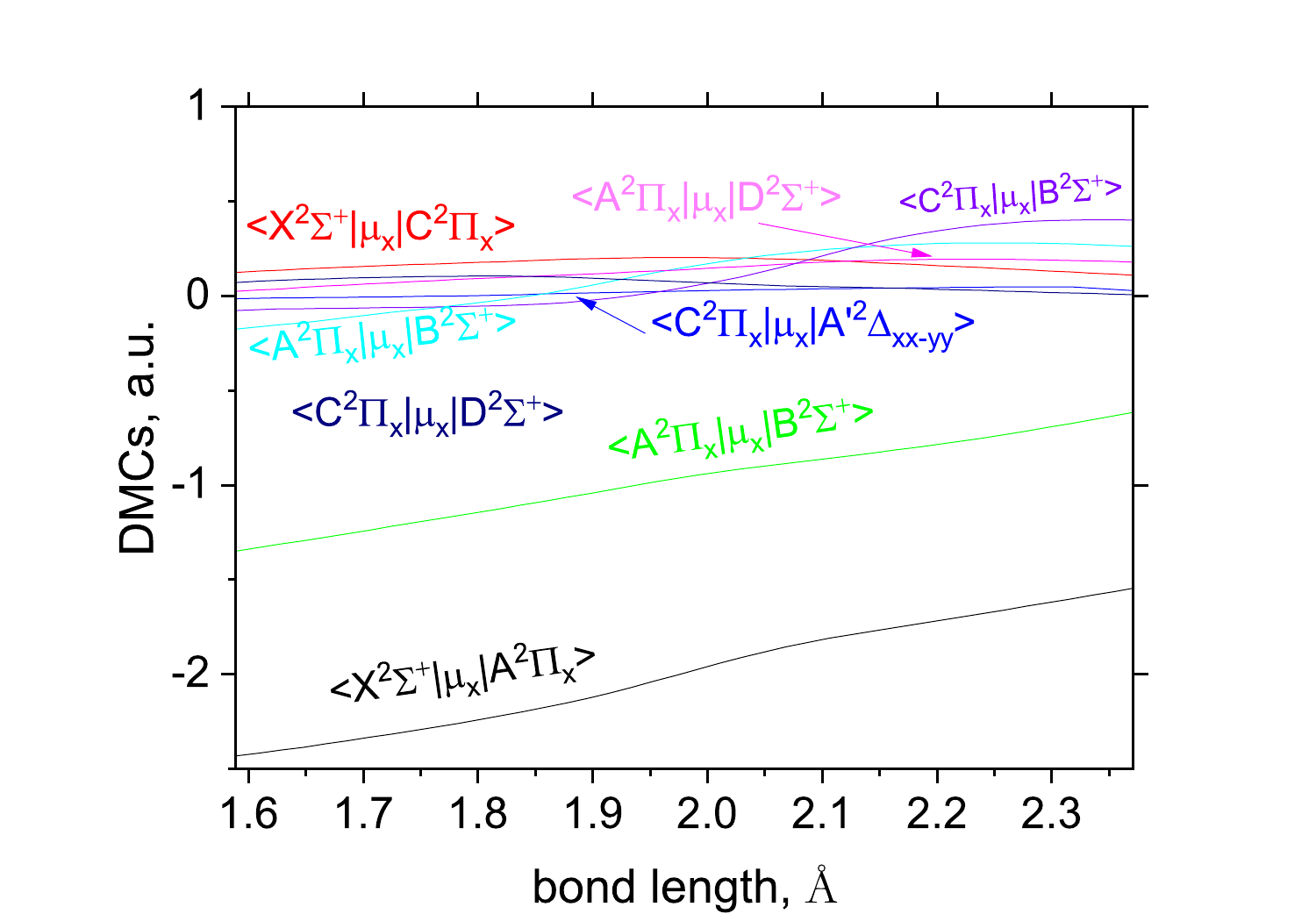}
    \caption{Diabatised \ai\ dipole moment matrix elements (in a.u.) as a function of bond length. The middle panel gives diagonal dipoles while
    the top and bottom panels give transition dipole moments.}
    \label{f:Dipoles}
\end{figure}

\begin{figure}
    \centering
    \includegraphics[width=\linewidth]{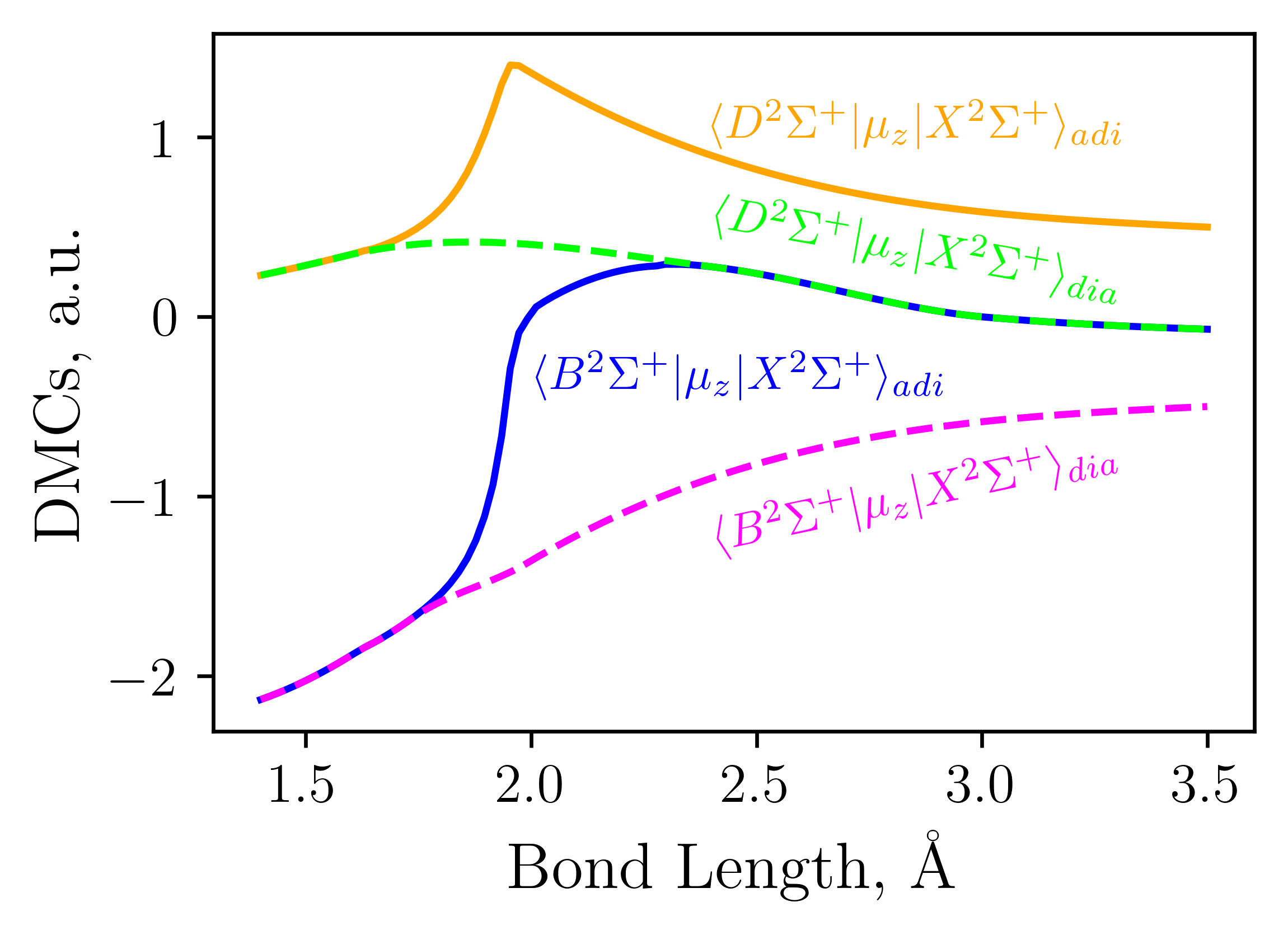}
    \caption{Example of the diabatisation (see text for details) of the adiabatic $\bra{\BSM}\mu_z\ket{ \XSM}$ and $\bra{\DSM}\mu_z\ket{\XSM}$ dipole moment pairs, where the \BS\ and \DS\ states exhibit an avoided crossing at $r \sim 1.81$~\AA.}
    \label{fig:dipole_diab_example}
\end{figure}

We diabatise our DMCs using a combination of cubic-spline interpolation to smooth out the region around the avoided crossing and knowledge of the shape of  the diabatised target curves.  Figure \ref{fig:dipole_diab_example} illustrates our property-based diabatising `transformation' for the $\bra{\BSM}\mu_z\ket{\XSM}$ and $\bra{\DSM}\mu_z\ket{\XSM}$  transition dipole moment pairs, the effect being the two curves `swap' beyond the avoided crossing and are now smooth.
Figure \ref{f:Dipoles} shows all  diabatised \ai\ diagonal and off-diagonal DMCs, which are smooth over all bond lengths.

Within nuclear motion and intensity calculations dipoles are sometimes represented as a grid of \ai\ points, however, one sees a flattening of the ground state IR overtone bands  with vibrational excitation. The source of this nonphysical flattening has been discussed by \citet{15MeMeSt.CO,16MeMeSt,22MeUs.CO,23UsSeYu}. It comes from numerical noise in the calculations, which is enhanced by the interpolation of the given \molpro\ dipole grid points onto the \duo\ defined grid.  The most effective method to reduce this numerical noise is  to represent the input dipole moments analytically \citep{15MeMeSt.CO}. We chose to represent our \XS\ DMC using the `irregular DMC' proposed by \citet{22MeUs.CO} which takes the form
\begin{equation}
\label{eq:irreg_cheby}
    \mu_{\rm irreg}(r)= \chi(r;c_2,...,c_6)\sum_{i=0}^6 b_iT_i(z(r))
\end{equation}
where $T_i$ are Chebyshev polynomials of the first kind, $b_i$ are summation coefficients to be fitted, $z(r)$ is a reduced variable in bond length and is given by
\begin{equation}
    z(r)=1-2e^{-c_1r}
\end{equation}
which maps the $r\in[0,\infty]$~\AA\ interval to the $z\in[-1,1]$ reduced interval (the region in which the Chebyshev polynomials have zeros), and finally $\chi(r;c_2,...,c_6)$ is an $r$-dependent term parametrically dependent on 5 $c_k$ parameters to be fitted and is given by
\begin{equation*}
\chi(r;c_2,...,c_6)\frac{(1-e^{-c_2r})^3}{\sqrt{(r^2-c_3^2)^2+c_4^2}\sqrt{(r^2-c_5^2)^2+c_6^2}}.
\end{equation*}

\begin{figure}
    \centering
    \includegraphics[width=\linewidth]{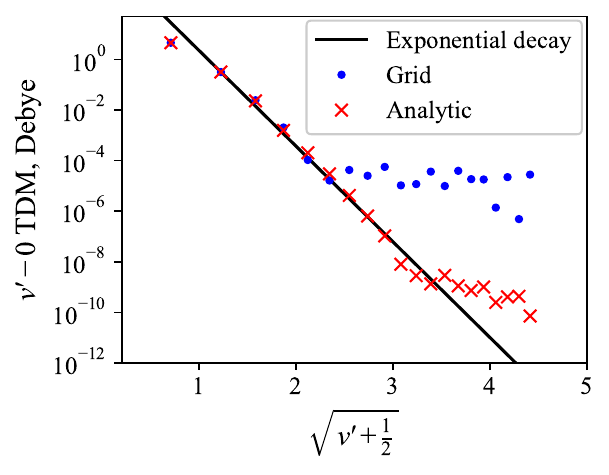}
    \caption{\XS\ $\rightarrow$ \XS\ $v'-0$ overtone TDMs are plotted on a log scale vs. $\sqrt{v'+\frac{1}{2}}$ and are computed using the grid interpolated \ai\ DMC (shown as blue circles) and our fitted analytical model DMC (Eq. (\ref{eq:irreg_cheby}), shown as red crosses). A simple exponential decay is shown for comparison which simulates the expected  NIDL-like behaviour \citep{12Medvedev}.}
    \label{fig:overtone_TDM}
\end{figure}

The irregular DMC form has the desirable properties of quickly converging to the correct long-range limit, having enough parameters (13) to ensure minimal local oscillations, and provide a straight Normal Intensity Distribution Law (NIDL) \citep{22MeUs.CO,12Medvedev,15MeMeSt.CO}.
This straight NIDL is a major restriction to the model DMC and means the logarithm of the overtone vibrational  transition dipole moments (VTDM) $\bra{v'}\mu(r)\ket{v=0}$ ($v'>1$) should evolve linearly with the square root of the upper state energy over the harmonic frequency, or $\sqrt{v'+\frac{1}{2}}$. Here we compute VTDMs  $\bra{v'}\mu(r)\ket{v=0}$ up to dissociation for the \XS\ using both the grid-defined dipole and the fitted analytical form, where figure \ref{fig:overtone_TDM} shows their behaviour. The expected linear behaviour of the NIDL is shown in Fig.~\ref{fig:overtone_TDM} as a gray line which is seen to  better agree with the TDM computed using the analytical \XS\ DMC compared to the calculation using the grid-interpolated DMC. At the $v'=6$ overtone the grid-interpolated DMC causes a non-physical flattening of the VTDM at $\sim 3.4\times 10^{-5}$ Debye, whereas we only see a departure from the straight NIDL at $v'=15$ when using the analytical form which flattens at $\sim 7.4\times10^{-10}$ Debye. The analytically represented \XS\ DMC therefore provides a more physically meaningful behaviour of the vibrational overtone VTDM but still departs from the expected NIDL at high overtones where the intensities are much lower and therefore less important.

Following \citet{19SmSoYu}, we scaled the \ai\ DMC of \XS\  by the factor 1.025 to match the experimental  value of the equilibrium dipole $\bra{X},v=0|\mu(r)|\ket{X,v=0}$ determined by \citet{90SuLoFr.YO}. The DMCs of \ApS, \AS\ and \BS\ were scaled by 0.97,   0.86 and 0.6, respectively to match the more accurate CCSD(T)/CBS single point calculations from \citet{19SmSoYu}.

The  \AS--\XS, \BS--\XS\ and \DS--\XS\ TDMCs had to be scaled  by 0.8, 0.75 and 2.8, respectively, to improve the agreement of the  corresponding calculated values of the \AS, \BS\ and \DS\ lifetimes  with the measurements of \citet{77LiPaxx.YO} and \citet{17ZhZhZh.YO}, see discussion below.



\section{Refinement of the spectroscopic model}

We use the diatomic code \Duo\ \citep{Duo}  to solve a  coupled system of Schr\"{o}dinger equations. \Duo\ is a free-access rovibronic solver for diatomic molecules available at \href{https://github.com/exomol/Duo/}{https://github.com/exomol/Duo/}. The hyperfine structure was  ignored.   For nuclear motion calculations a vibrational sinc-DVR basis set was defined as a grid of 151 internuclear geometries in the range of {1.4}--{3.5}~{\AA}. We select the lowest 30, 30, 35, 30, 30, 30 vibrational wavefunctions of the \XS, \ApS, \AS, \BS, \CS, and \DS\ states, respectively, to form the contracted vibronic basis. A refined spectroscopic model of YO was obtained by fitting the expansion parameters representing different properties to 5089 empirically derived rovibrational energy term values of $^{89}$Y$^{16}$O described above.

\begin{figure*}
    \centering
    \includegraphics[width=0.44\textwidth]{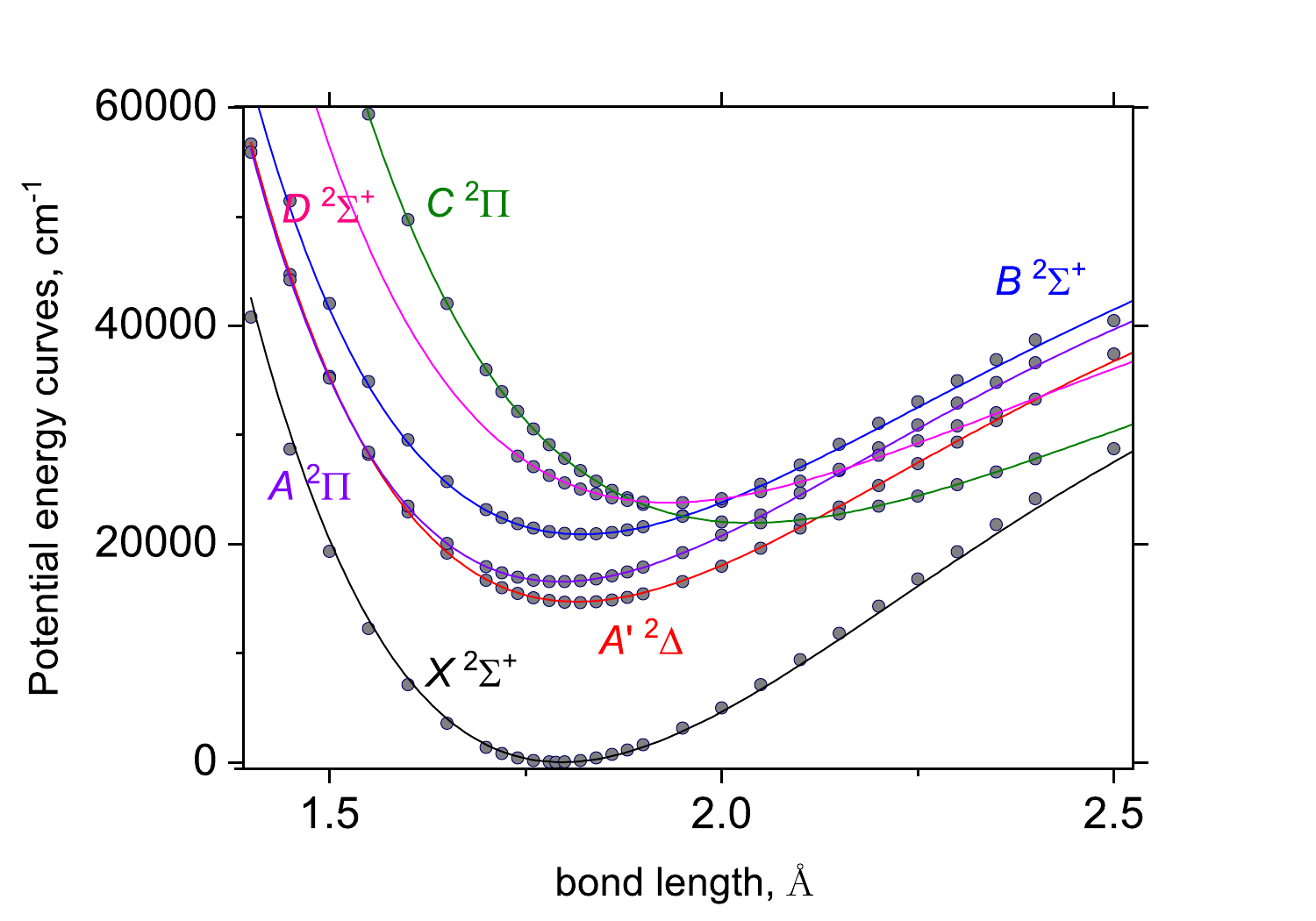}
    \includegraphics[width=0.44\textwidth]{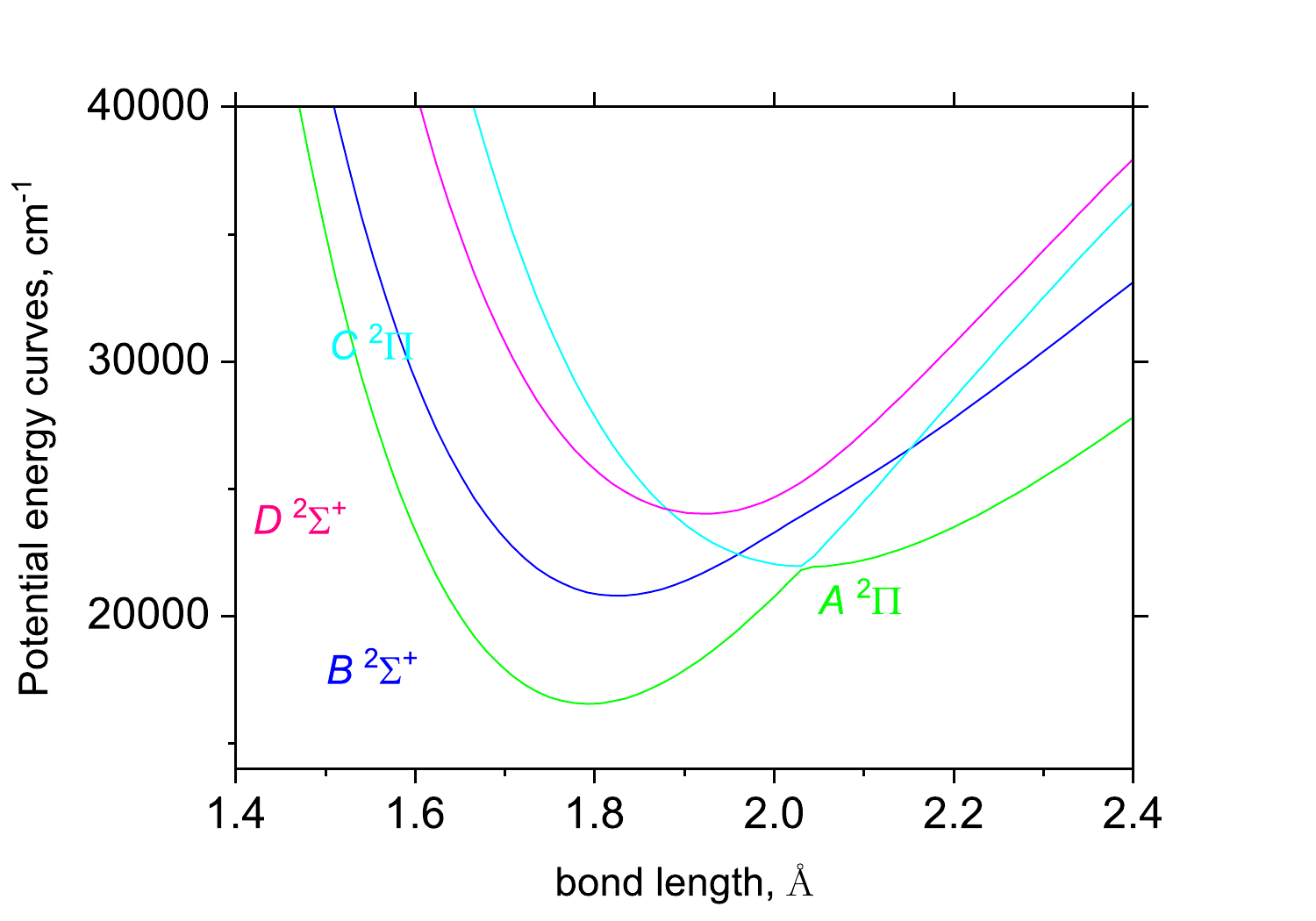}
\caption{Refined (lines) and \ai\ (points) PECs of YO:  diabatic (left) and adiabatic (right). }
    \label{f:PECS:dia}
\end{figure*}

The refined (diabatic) PECs of YO are illustrated in Fig.~\ref{f:PECS:dia}. The CCSD(T)/CBS \ai\ energies from \citet{19SmSoYu}, shown with circles, appear to closely follow the refined curves, indicating the excellent quality of the \ai\ CCSD(T) PECs.

\begin{figure*}
    \centering
    \includegraphics[width=0.44\textwidth]{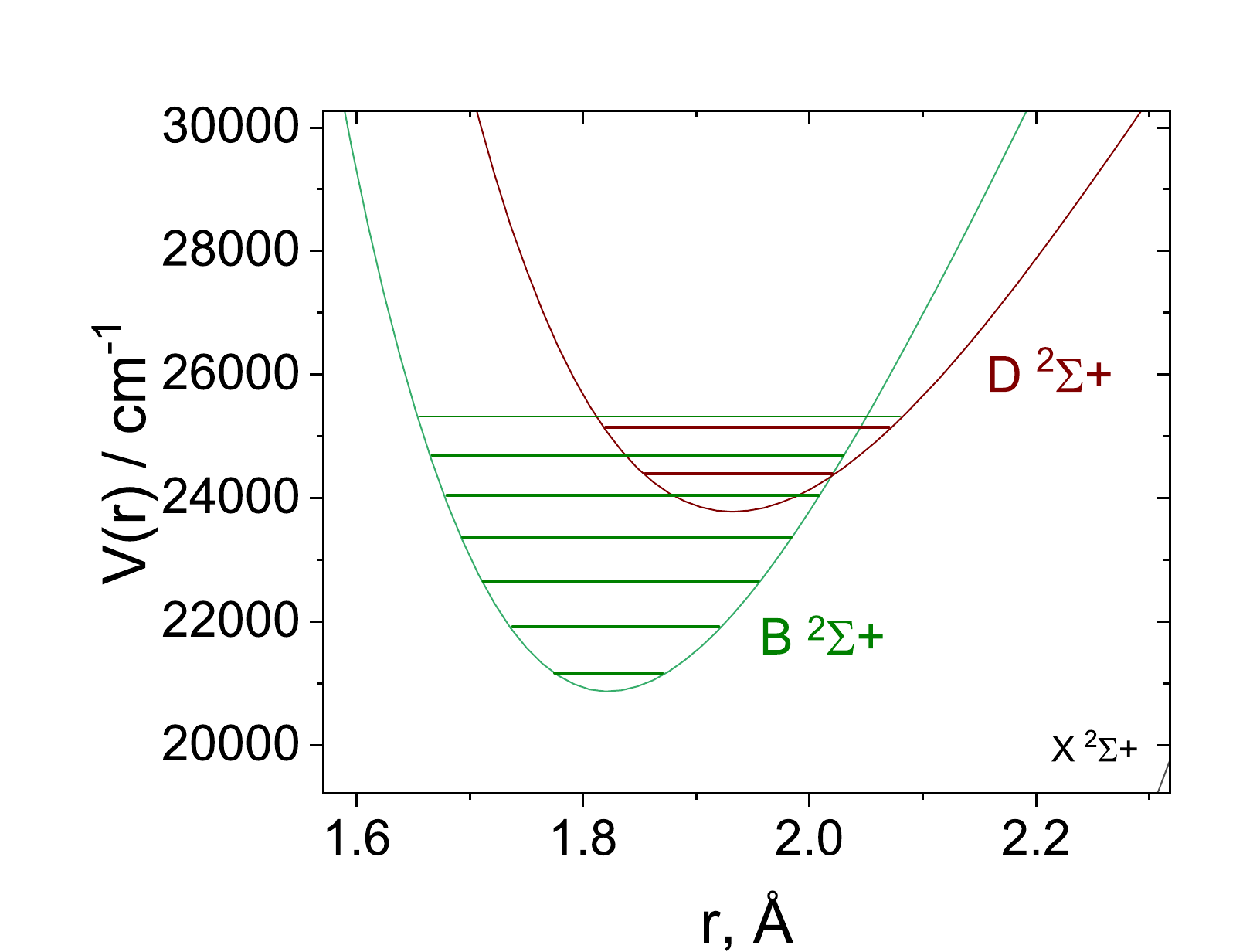}
    \includegraphics[width=0.44\textwidth]{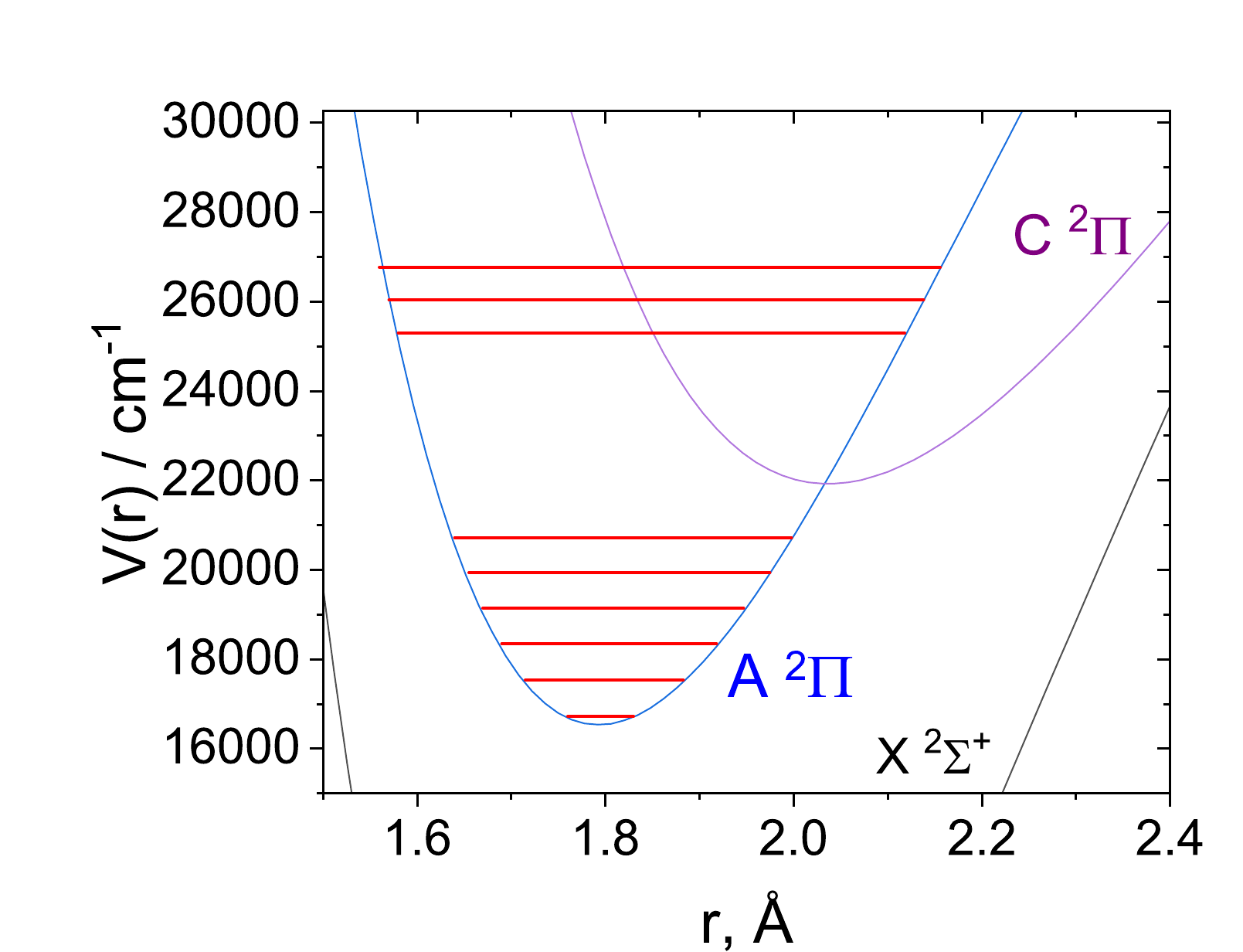}
\caption{Diabatic PECs of the $B$/$D$ and $A$/$C$ pairs with the corresponding experimental energy term values ($J=0.5$).}
    \label{f:energies:B:D:A:C}
\end{figure*}

Diabatic representations of the \AS, \BS, \CS\ and \DS\  states   are illustrated  in Fig.~\ref{f:energies:B:D:A:C}, where the corresponding experimental  energy term values are also shown  ($J=0.5$ or $J=1.5$).
Due to the very close positioning of the \DS\ ($v=2$) and \BS\ ($v=6$) states,  the \BS\ ($v=6$) rovibronic wavefunctions appear strongly mixed with the \DS\ ($v=2$) wavefunctions in the  \Duo\ solution, especially at  higher $J$.

The \BS\ vibronic energies of $v\ge 4$ are strongly affected by the diabatic coupling with the \DS\ state. Introduction of the diabatic coupling to the \BS\ states makes  the shape of the PEC broader and pushes  the positions of the \BS\ energies down. It is interesting to note that  the \AS\ state vibronic energies  for $v=11, 12, 13$ do not appear to be very perturbed by  the presence of the close-by \CS\ state, unlike the interaction of the $B$/$D$ diabatic pair.
This can be attributed to the difference in the corresponding NACs of the $B$/$D$ and $A$/$C$ pairs in Fig.~\ref{f:NAC}.

By construction, all \Duo\ eigenfunctions and eigenvalues are automatically assigned the rigorous quantum numbers $J$ and parity $\tau = \pm 1$. To assign the non-rigorous rovibronic quantum numbers,  \Duo\ first defines the spin-electronic components (`State' and $\Omega$)  using the largest contribution from the eigen-coefficients approach \citep{Duo}. Within each rotation-spin-electronic state, the vibrational excitation is then defined by a simple count of the increasing energies starting from $v=0$.

The refined SOCs, EAMCs, TDMCs of YO are shown in Fig.~\ref{f:SOCs}. The refined diabatic couplings for the $B$--$D$ and $A$--$C$ pairs are shown in Fig.~\ref{f:NAC}.

All parameters defining the final  spectroscopic model of YO are included in the supplementary material as a \Duo\ input file.

\begin{figure*}
    \centering
    \includegraphics[width=0.44\textwidth]{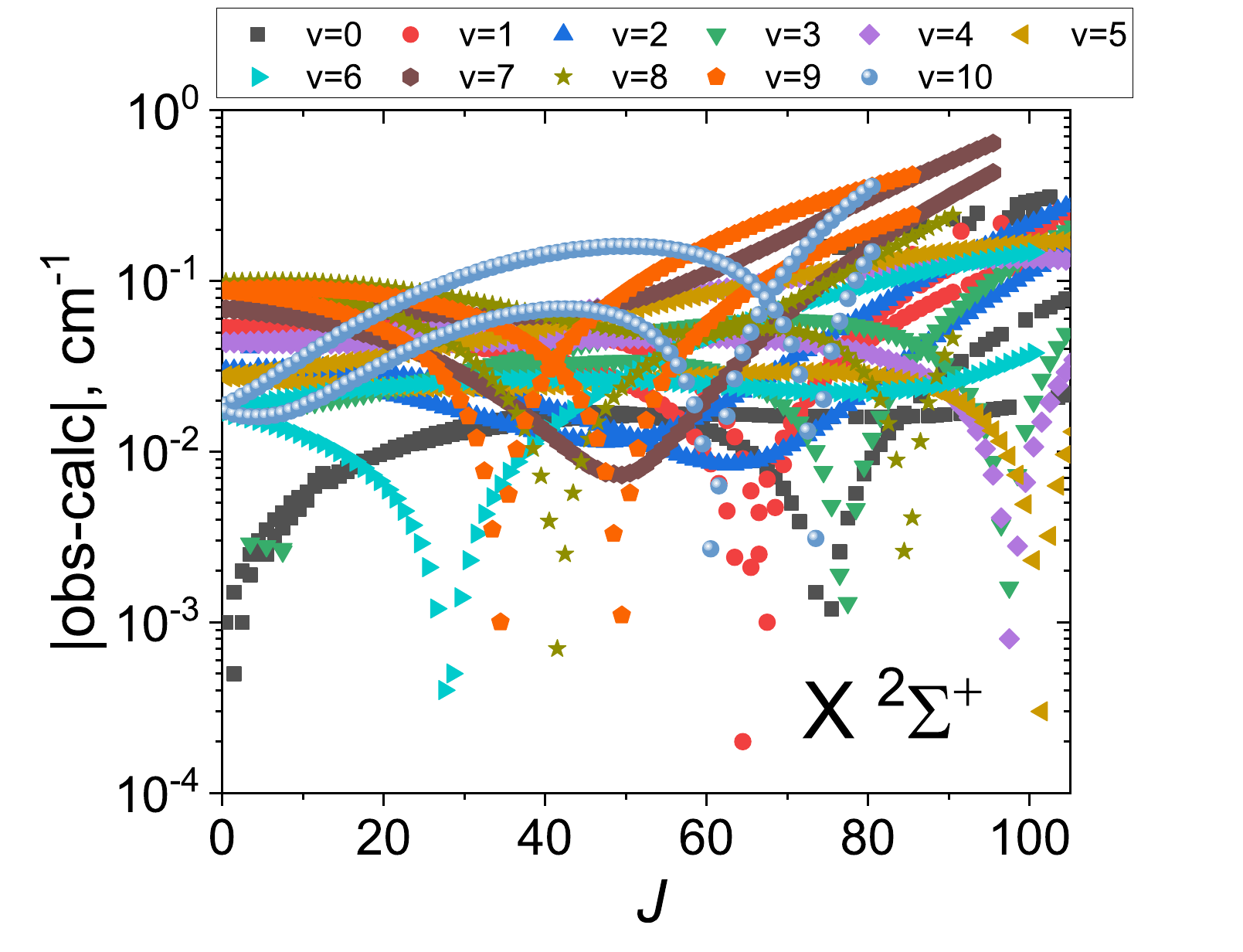}
    \includegraphics[width=0.44\textwidth]{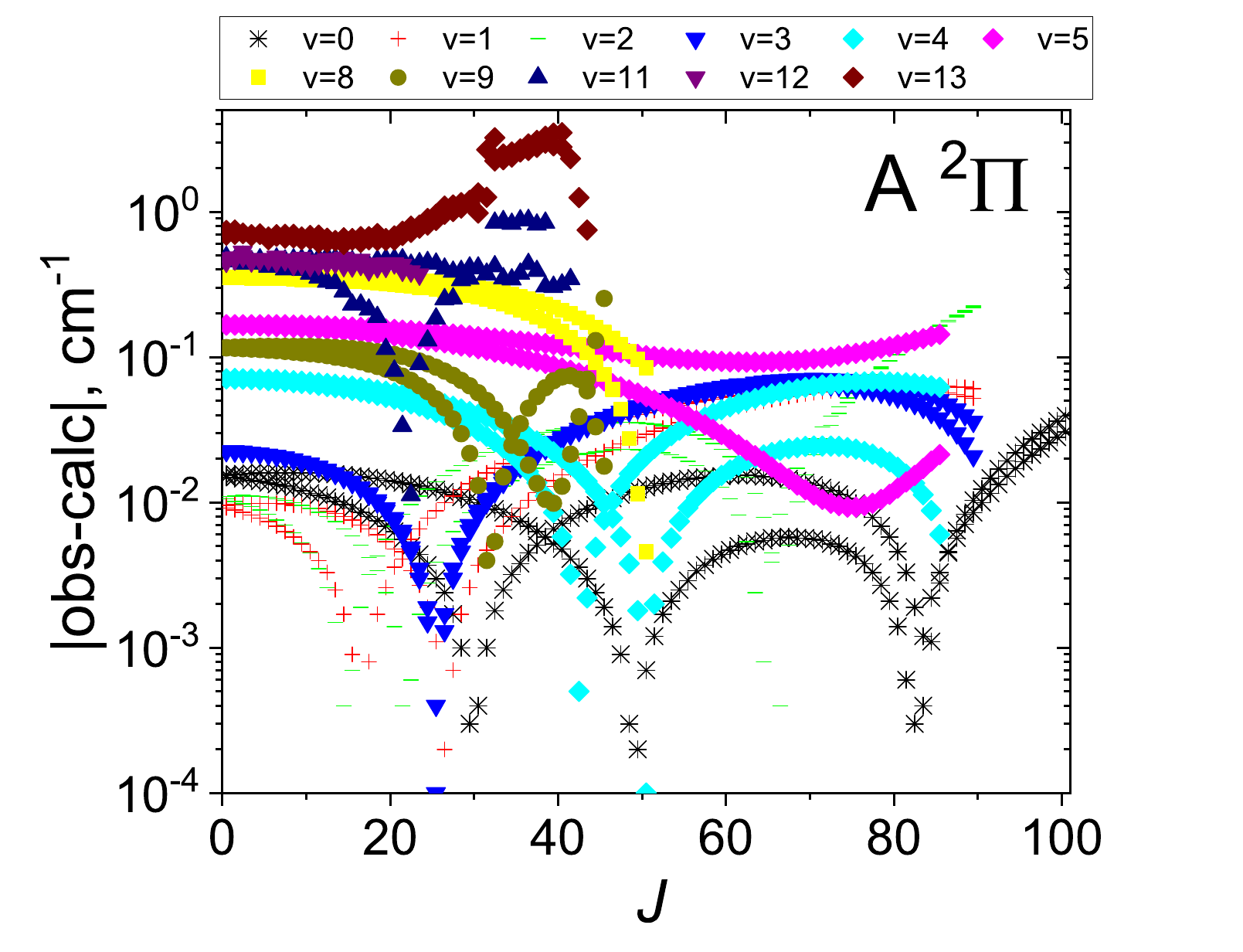}
    \includegraphics[width=0.44\textwidth]{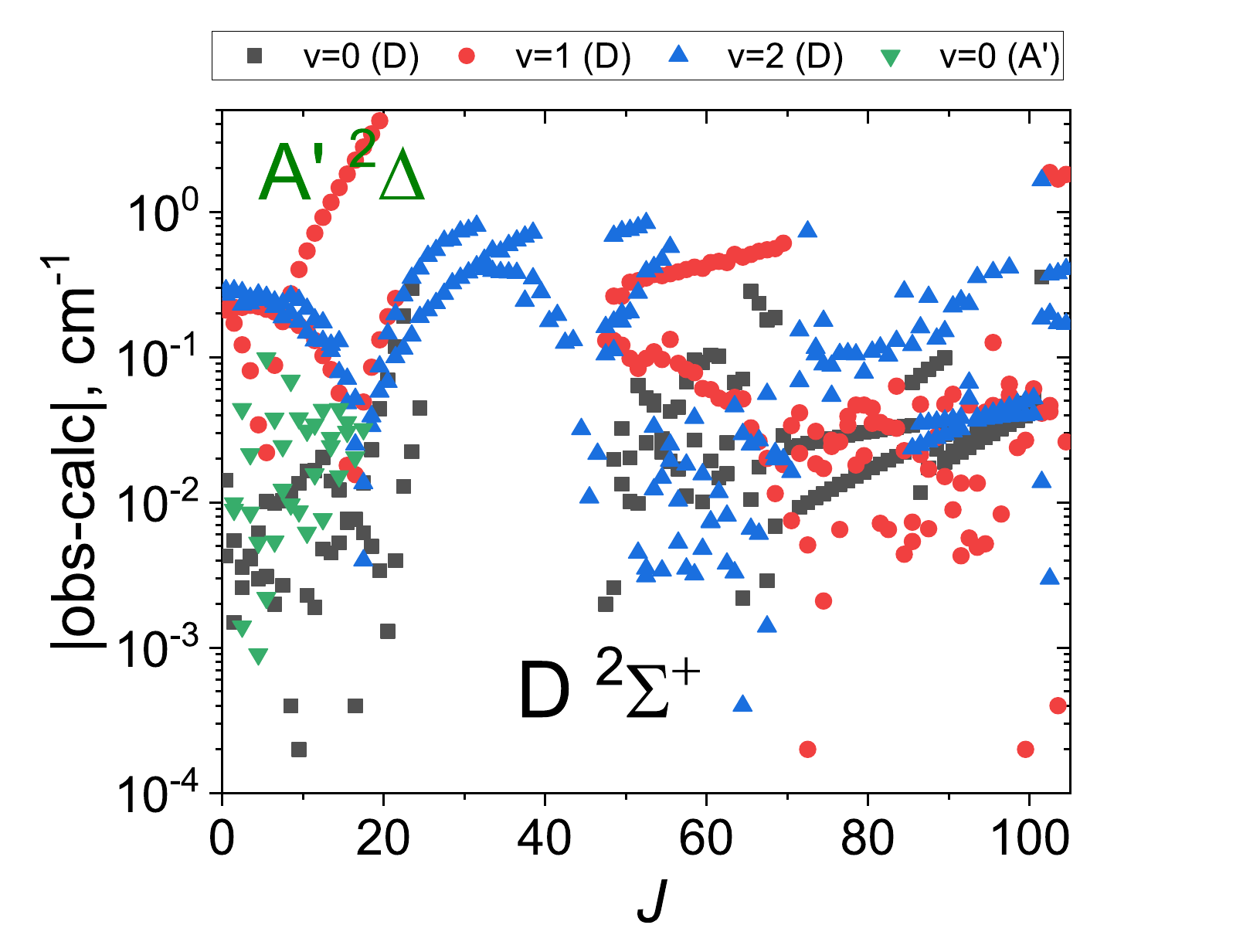}
    \includegraphics[width=0.44\textwidth]{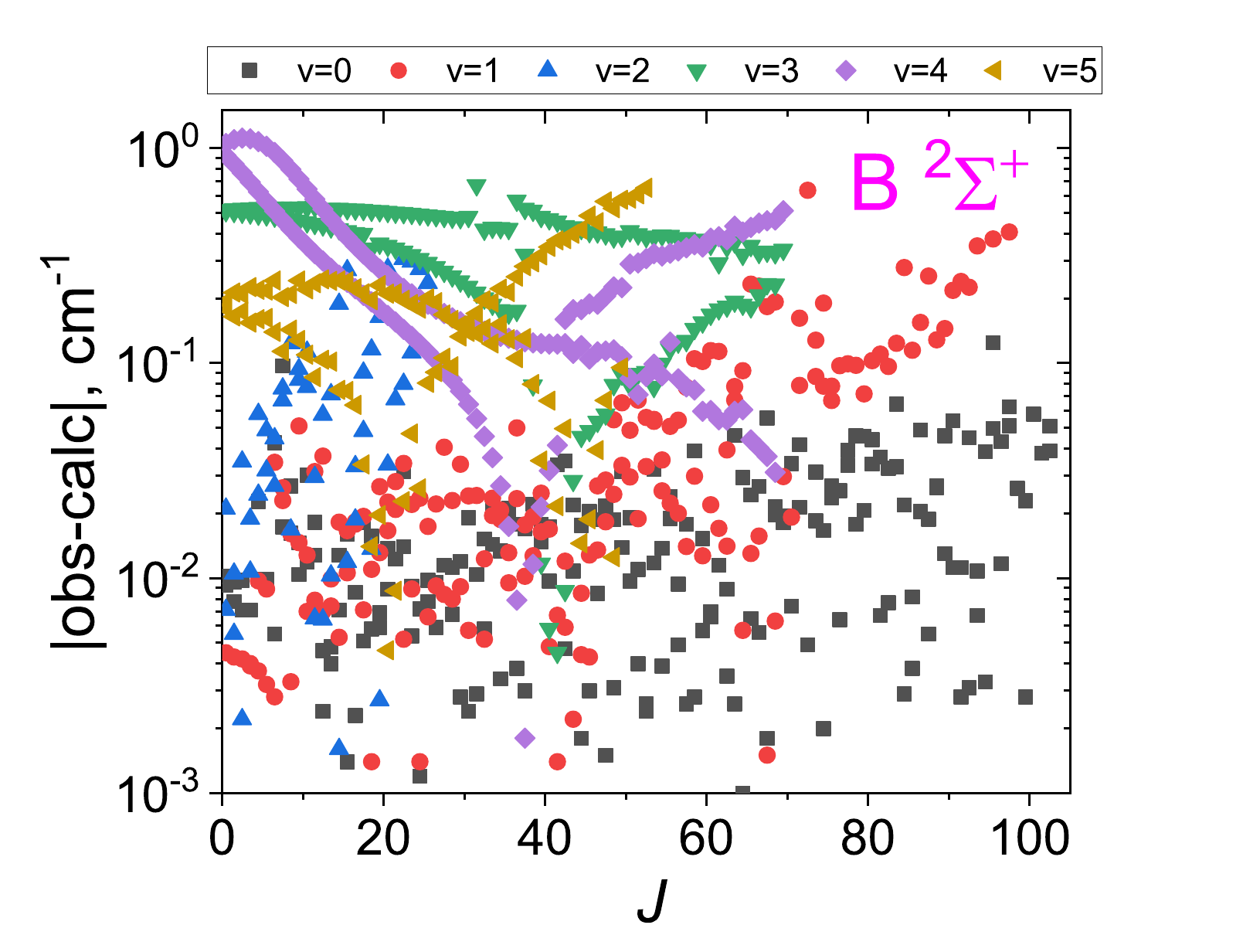}

\caption{Observed minus calculated residuals for YO using the refined spectroscopic model for different vibronic states.}
    \label{f:obs-calc}
\end{figure*}

The results of the fittings are illustrated in Fig.~\ref{f:obs-calc}, where $|$obs.-calc.$|$ residuals are shown for different electronic states. Some of the bands show clear systematic behavior of the residuals with respect to $J$, especially those that correspond to the synthetic data or high resolution data (e.g. \XS, \AS\ and some of the \BS\ systems), while others appear random with no particular structure  (e.g. $v=0$ of \BS\ and \DS). No other obvious trends could be seen.

The root-mean-squares error achieved is 0.29~\cm\ for all 5906  energies covering  $J$ up to 142.5.


There are no experimental data on the \CS\ state due to its large displacement from the Franck-Condon region of the \XS\ state. We therefore had to rely on the available \ai\ curves associated with this state as well as on their quality. Unlike the CCSD(T) PECs, the corresponding coupling curves were computed with MRCI and are less accurate.  Moreover, there is no experimental data representing  perturbations caused by the \CS\ rovibronic state on other vibronic states in the limited experimental data on YO. However theoretically, using the \ai\ data, we do see  such  perturbations in the \AS, \BS\ and \DS\ states due to the spin-orbit and EAM couplings with \CS\ (see \citet{19SmSoYu}), which makes the fit especially difficult. We therefore decided to switch off all the coupling with the \CS\ state in this work.

The experimental data on the  \ApS\ state is limited to  $v=0$ ($J \leq 17.5$), which means only the  potential minimum $V_{\rm e}$ of the \ApS\ state and the corresponding equilibrium constant could be usefully refined, but not its shape, which was fixed to the \ai\ CCSD(T) curve via the corresponding EHH potential parameters.

\section{Line List}
\label{s: Line List}

Using our final semi-empirical spectroscopic model, a rovibronic line list of $^{89}$Y$^{16}$O  called \name\ covering  the lowest 6 doublet electronic states and the wavelength range up to 166.67~nm was produced.  In total 60~678~140 Einstein $A$ coefficients between 173~621 bound rovibronic states were computed with a maximum total rotational quantum number $J_\text{max}$ = 400.5. $^{89}$Y is the only stable isotope of yttrium; however, using the same model, line lists for two minor isotopologues $^{89}$Y$^{17}$O  and $^{89}$Y$^{18}$O  have been also generated.

The line lists are presented in the standard ExoMol format \citep{jt548,jt810} consisting of a States file and a Transitions file with extracts shown in Tables~\ref{t:states} and \ref{t:trans}, respectively. The calculated energies in the States file  are `MARVELised', i.e. we replace  them  with the (pseudo-)MARVEL values where available. The uncertainties are taken as the experimental (pseudo-)MARVEL uncertainties for the substituted values,  otherwise the following empirical and rather conservative expression is used:
\begin{equation}
\label{e:unc:form}
{\rm unc.} = \Delta T + \Delta \omega\, v + \Delta B\,  J (J+1),
\end{equation}
with the state-dependent parameters listed in Table~\ref{t:unc:const}.

\begin{figure}
	\includegraphics[width=0.5\textwidth]{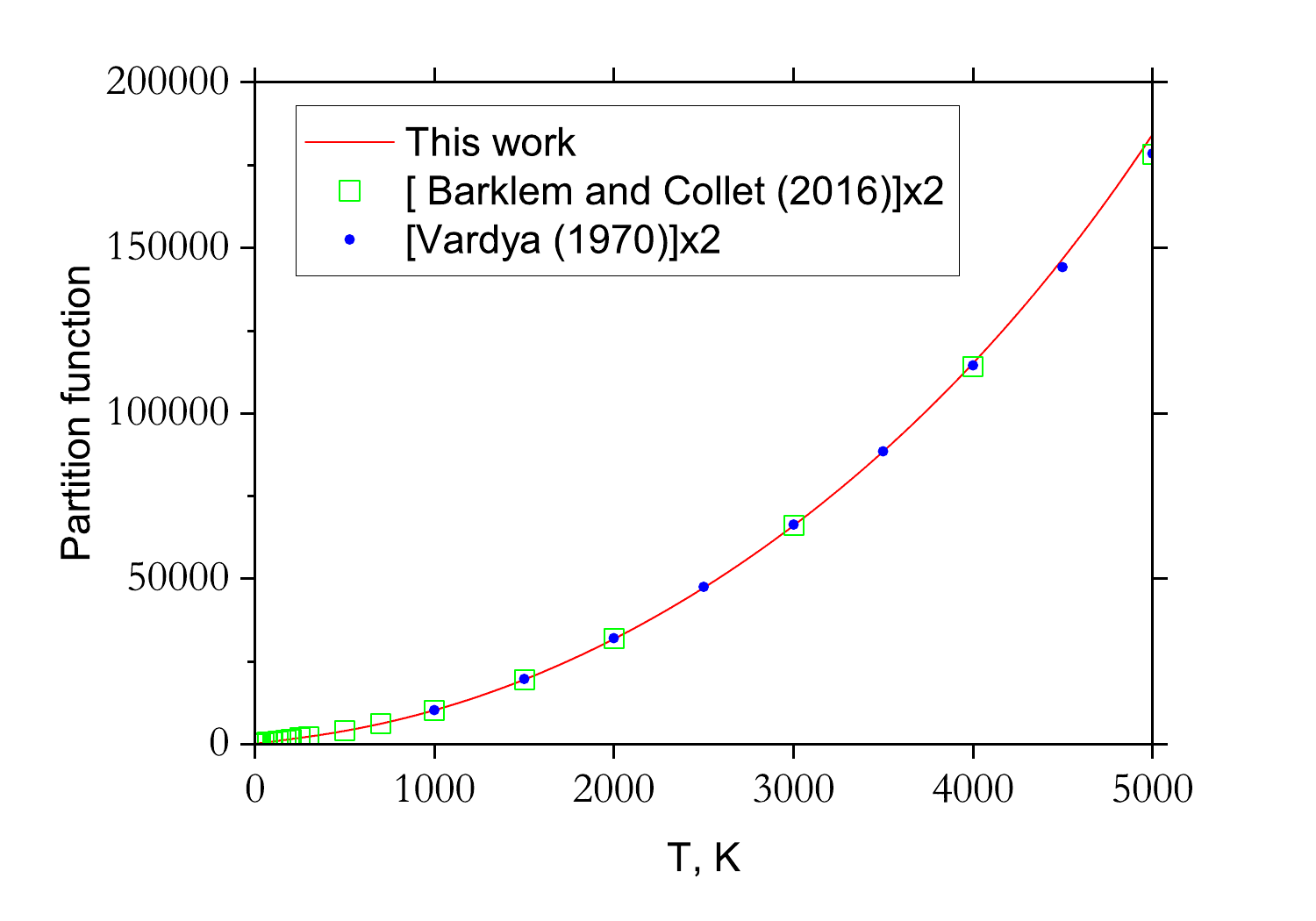}
	\caption{Partition functions of YO: From this work (solid line), from \protect\citet{70Vardya.YO}  (filled circles) and from \protect\citet{16BaCoxx.partfunc} (open squares). The latter two were multiplied by a factor of 2 to account for the different treatment of nuclear statistics. }
	\label{f:pf}
\end{figure}

The partition function of YO computed using the new line list is shown in Fig.~\ref{f:pf},  where it is compared to the partition functions by  \citet{16BaCoxx.partfunc} and  \citet{70Vardya.YO}, showing close agreement once a
correction is made for the fact that $^{89}$Y has nuclear spin $\frac{1}{2}$.
We also generate temperature- and pressure-dependent opacities of YO  using the \name\ line list and by following  the ExoMolOP procedure \citep{jt801} for four exoplanet atmosphere retrieval codes ARCiS \citep{ARCiS}, TauREx \citep{TauRex3}, NEMESIS \citep{NEMESIS} and petitRADTRANS \citep{19MoWaBo.petitRADTRANS}.

The \name\ line lists, partition function and opacities are available at \href{www.exomol.com}{www.exomol.com}.

\begin{table*}
\centering
\caption{ Extract from the states file of the line list for  YO.}
\label{t:states}
{\tt  \begin{tabular}{rrrrrrcclrrrrcrr} \hline \hline
$i$ & Energy (\cm) & $g_i$ & $J$ & unc &  $\tau$ & $g$&  \multicolumn{2}{c}{Parity} 	& State	& $v$	&${\Lambda}$ &	${\Sigma}$ & $\Omega$ & Ma/Ca & Energy (\cm) \\
\hline
     329 &   17109.384230 &    8 &   1.5 &   0.060200 &  0.000000   &  -0.000268  &   +   &   f   & A2Pi     &   1   &   1   & -0.5  &  0.5  &  Ma   &  17109.393106    \\
     330 &   17486.768827 &    8 &   1.5 &   0.220200 &  0.002928   &  -0.400448  &   +   &   f   & X2Sigma+ &  22   &   0   &  0.5  &  0.5  &  Ca   &  17486.768827    \\
     331 &   17538.074400 &    8 &   1.5 &   0.060200 &  0.000000   &  0.798897   &   +   &   f   & A2Pi     &   1   &   1   &  0.5  &  1.5  &  Ma   &  17538.062594    \\
     332 &   17635.333239 &    8 &   1.5 &   4.030000 &  0.000757   &  0.399545   &   +   &   f   & Ap2Delta &   4   &   2   & -0.5  &  1.5  &  Ca   &  17635.333239    \\
     333 &   17916.112120 &    8 &   1.5 &   0.110200 &  0.000000   &  -0.000268  &   +   &   f   & A2Pi     &   2   &   1   & -0.5  &  0.5  &  Ma   &  17916.122345    \\
     334 &   18210.832409 &    8 &   1.5 &   0.230200 &  0.002827   &  -0.400448  &   +   &   f   & X2Sigma+ &  23   &   0   &  0.5  &  0.5  &  Ca   &  18210.832409    \\
     335 &   18345.385550 &    8 &   1.5 &   0.110200 &  0.000000   &  0.798904   &   +   &   f   & A2Pi     &   2   &   1   &  0.5  &  1.5  &  Ma   &  18345.398884    \\
     336 &   18403.241526 &    8 &   1.5 &   5.030000 &  0.000554   &  0.399553   &   +   &   f   & Ap2Delta &   5   &   2   & -0.5  &  1.5  &  Ca   &  18403.241526    \\
\hline
\hline
\end{tabular}}
\mbox{}\\
{\flushleft
$i$:   State counting number.     \\
$\tilde{E}$: State energy term values in \cm, MARVEL or Calculated (\textsc{Duo}). \\
$g_i$:  Total statistical weight, equal to ${g_{\rm ns}(2J + 1)}$.     \\
$J$: Total angular momentum.\\
unc: Uncertainty, \cm.\\
$\tau$: Lifetime (s$^{-1}$).\\
$g$: Land\'{e} $g$-factors \citep{16SeYuTe}. \\
$+/-$:   Total parity. \\
$e/f$:   Rotationless parity. \\
State: Electronic state.\\
$v$:   State vibrational quantum number. \\
$\Lambda$:  Projection of the electronic angular momentum. \\
$\Sigma$:   Projection of the electronic spin. \\
$\Omega$:   Projection of the total angular momentum, $\Omega=\Lambda+\Sigma$. \\
Label: `Ma' is for MARVEL and `Ca' is for Calculated. \\
Energy: State energy term values in \cm, Calculated (\textsc{Duo}). \\
}
\end{table*}

\begin{table}
\centering
\caption{Extract from the transitions file of the line list for  YO. }
\tt
\label{t:trans}
\centering
\begin{tabular}{rrrr} \hline\hline
\multicolumn{1}{c}{$f$}	&	\multicolumn{1}{c}{$i$}	& \multicolumn{1}{c}{$A_{fi}$ (s$^{-1}$)}	&\multicolumn{1}{c}{$\tilde{\nu}_{fi}$} \\ \hline
   78884  &     78556  &        4.1238E+02  &    10000.000225   \\
  111986  &    112128  &        6.0427E+01  &    10000.000489   \\
   69517  &     69133  &        5.2158E-03  &    10000.000708   \\
   39812  &     40514  &        7.3060E-02  &    10000.000818   \\
   34753  &     33815  &        6.2941E-04  &    10000.001400   \\
   72754  &     72910  &        2.5370E+00  &    10000.001707   \\
  130747  &    130937  &        5.3843E-04  &    10000.002153   \\
  130428  &    130122  &        1.5407E-02  &    10000.002287   \\
  114934  &    115604  &        3.2160E+00  &    10000.002360   \\
   12357  &     11958  &        6.2849E+02  &    10000.002755   \\
  135752  &    135933  &        1.4867E+01  &    10000.004338   \\
    \hline\hline
\end{tabular} \\ \vspace{2mm}
\rm
\noindent
$f$: Upper  state counting number;\\
$i$:  Lower  state counting number; \\
$A_{fi}$:  Einstein-$A$ coefficient in s$^{-1}$; \\
$\tilde{\nu}_{fi}$: transition wavenumber in \cm.\\
\end{table}

\begin{table}
\centering
\caption{ Parameters used to estimate uncertainties of the YO calculated energy term values according to Eq.~\eqref{e:unc:form}.}
\label{t:unc:const}
\begin{tabular}{lrrrr}
\hline \hline
State & $\Delta T$ & $\Delta \omega$ & $\Delta B$  \\
\hline
\XS\ & 0.0	&0.01	&0.0001 \\
\AS\ & 0.01	&0.05	&0.0001 \\
\CS\ & 10.0	&1.0	&0.01  \\
\BS\ & 0.01	&0.057&	0.0001 \\
\DS\ & 0.01	&0.07&	0.0001 \\
\ApS\ & 0.01	&1	&0.01 \\
    \hline\hline
\end{tabular}
\end{table}
\section{Simulated Spectra}
\label{sec: Simulated Spectra}

\begin{figure}
    \centering
    \includegraphics[width=\linewidth]{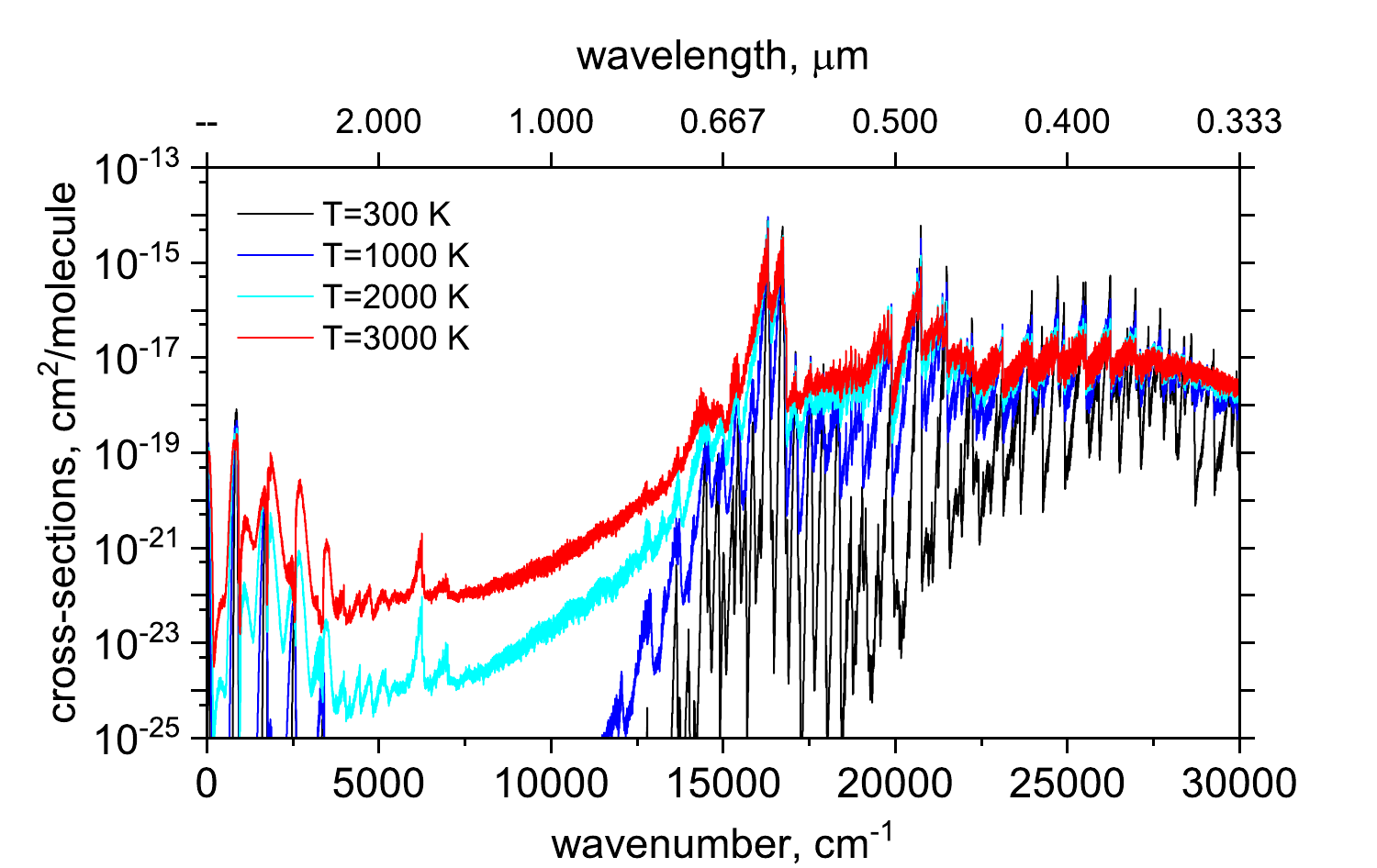}
    \caption{The simulated YO absorption spectrum computed at different temperatures. We adopt a Lorentzian line broadening of 1 \cm\ for each line which is computed at a resolution of 1 \cm. We see the intensity deviation is greatest around 0.5 and 0.6~$\mu$m  where the \XS$\rightarrow$\AS\ and \XS$\rightarrow$\BS\  bands  dominate opacity.}
    \label{fig:temp_spec}
\end{figure}

\begin{figure}
    \centering
    \includegraphics[width=\linewidth]{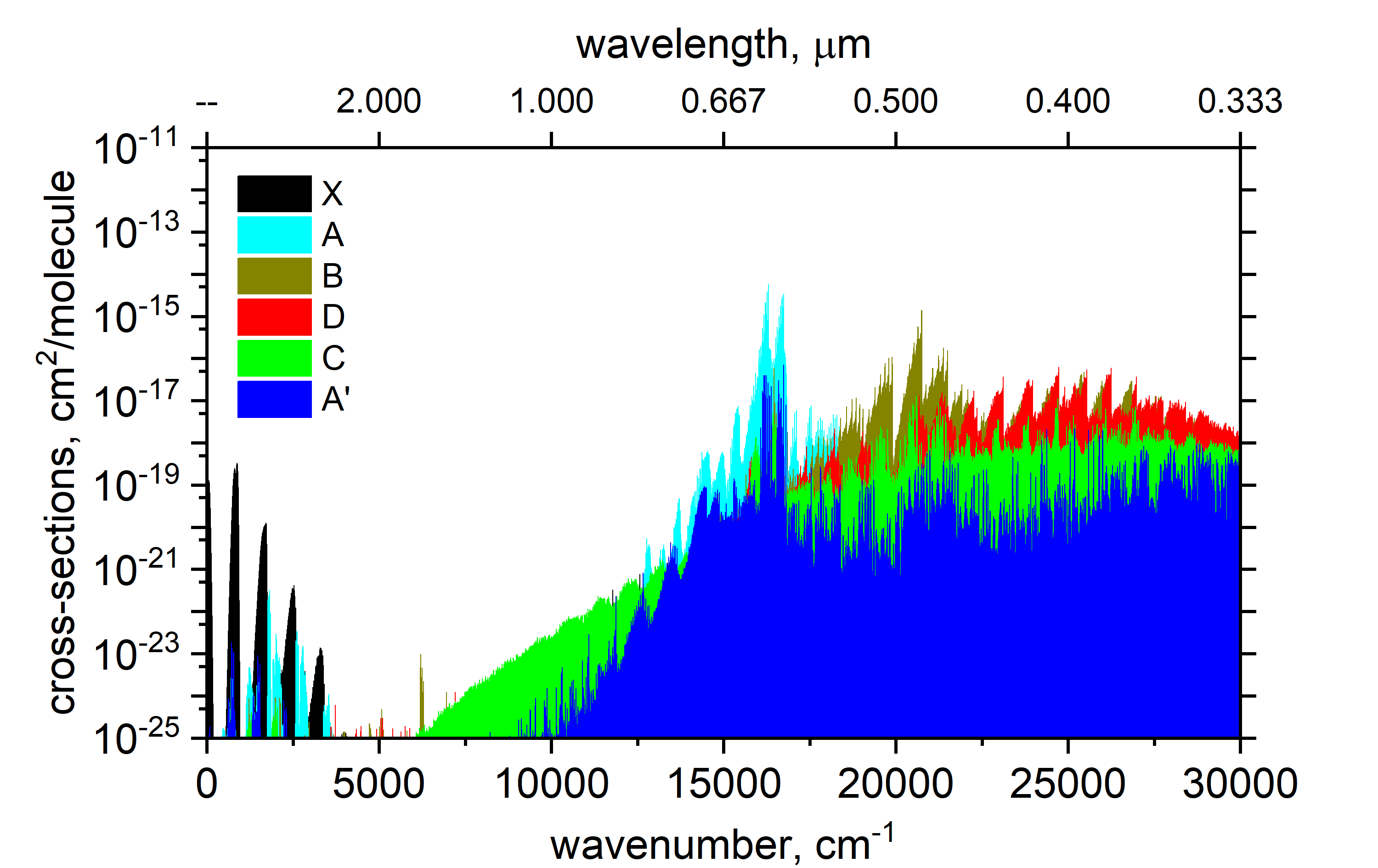}
    \caption{Different electronic band components of the absorption spectrum simulated at 2000 K using Lorentzian line broadening of 1 \cm\ for each line computed at a resolution of 1 \cm.}
    \label{fig:comp_spec}
\end{figure}

Using the computed line list for YO, here we simulate the YO rovibronic absorption spectra using the program \exocross\ \citep{ExoCross}.
Fig.~\ref{fig:temp_spec} shows the temperature variation of the YO rovibronic spectrum over the spectroscopic range up to 166 nm. Figure \ref{fig:comp_spec} highlights the contribution of the most important electronic bands to the total opacity simulated at 2000 K such that there is good separation between the electronic bands. In both simulations, each line was broadened using a Lorentzian line profile with a HWHM of 1 \cm\ and computed at a resolution of 1 \cm.


\subsection{Lifetimes}

Table~\ref{t:lf} compares our lifetimes to the experimental and theoretical values from the literature: laser  fluorescence measurements of the \AS\ and \BS\ states ($v\le 2$) of YO by \citet{77LiPaxx.YO} and \BS\ ($v=0$) and \DS\ ($v=0,1$)  lifetimes by \citet{17ZhZhZh.YO} as well as to the theoretical values by \citet{88LaBaxx.YO} and \citet{19SmSoYu}. The theoretical values correspond to the lowest $J$ values, $J=0.5$ for \XS, \BS\ and \DS, $J=1.5$ for \AS\ and \CS, $J=1.5$ for \ApS\, which we consider as a good proxy for the experimental values ($J$ unspecified) due to very slow $J$ dependence of the lifetimes. The good agreement is partly due to the adjustment of the corresponding TDMC to match the corresponding lifetimes as specified above. Our result is the best we could do for a complicated system \DS--\XS\ with a complex diabatic coupling (Fig.~\ref{f:NAC}) and the diabatised \DS--\XS\  TDMC based on some level of arbitrariness   (Fig.~\ref{fig:dipole_diab_example}).

\begin{table}
\caption{Lifetimes of $^{89}$Y$^{16}$O states in ns: comparison with the measurements of [77LiPa] \citep{77LiPaxx.YO} and [17ZhZhZh] \citep{17ZhZhZh.YO}, and the \ai\ calculations of [19SmSoYu] \citep{19SmSoYu}. } \label{t:lf}
\begin{tabular}{lrcrrr}
\hline
\hline
State       &$v$ &   [77LiPa]    & [17ZhZhZh]&  [19SmSoYu] &   This work  \\
\AS$_{1/2}$  &$      0  $&$     33.0 \pm 1. 3    $&$             $&$       22.6 $&$     35.5 $\\
             &$      1  $&$      36.5\pm 2.4     $&$             $&$       23.0 $&$     36.1 $\\
\AS$_{3/2}$  &$      0  $&$     32. 3\pm 0. 9    $&$             $&$       20.9 $&$     32.7 $\\
             &$      1  $&$       30.4\pm1.8     $&$             $&$       21.3 $&$     33.3 $\\
             &$      2  $&$     33.4 \pm 1. 5    $&$             $&$       21.6 $&$     33.9 $\\
             &$      6  $&$      41.6\pm 2.1     $&$             $&$       29.2 $&$     37.3 $\\
\BS\         &$      0  $&$      30.0\pm 0.9     $&$   38\pm 5   $&$       32.5 $&$     30.7 $\\
             &$      1  $&$     32.5 \pm 1. 2    $&$             $&$       34.3 $&$     30.4 $\\
\DS\         &$      0  $&$                      $&$   79 \pm 5  $&$       30.1 $&$     62.5 $\\
             &$      1  $&$                      $&$   79 \pm 5  $&$       29.2 $&$     56.8 $\\
             &$         $&$                      $&$             $&$            $&$          $\\
             &$         $&$                      $&$             $&$            $&$          $\\
             &$         $&$                      $&$             $&$            $&$          $\\
             &$         $&$                      $&$             $&$            $&$          $\\
\hline
\end{tabular}
\end{table}

\section{Comparisons To Experimental Spectra}
\label{sec: Comparisons to experimental Spectra}

\begin{figure*}
    \centering
    \includegraphics[width=\linewidth]{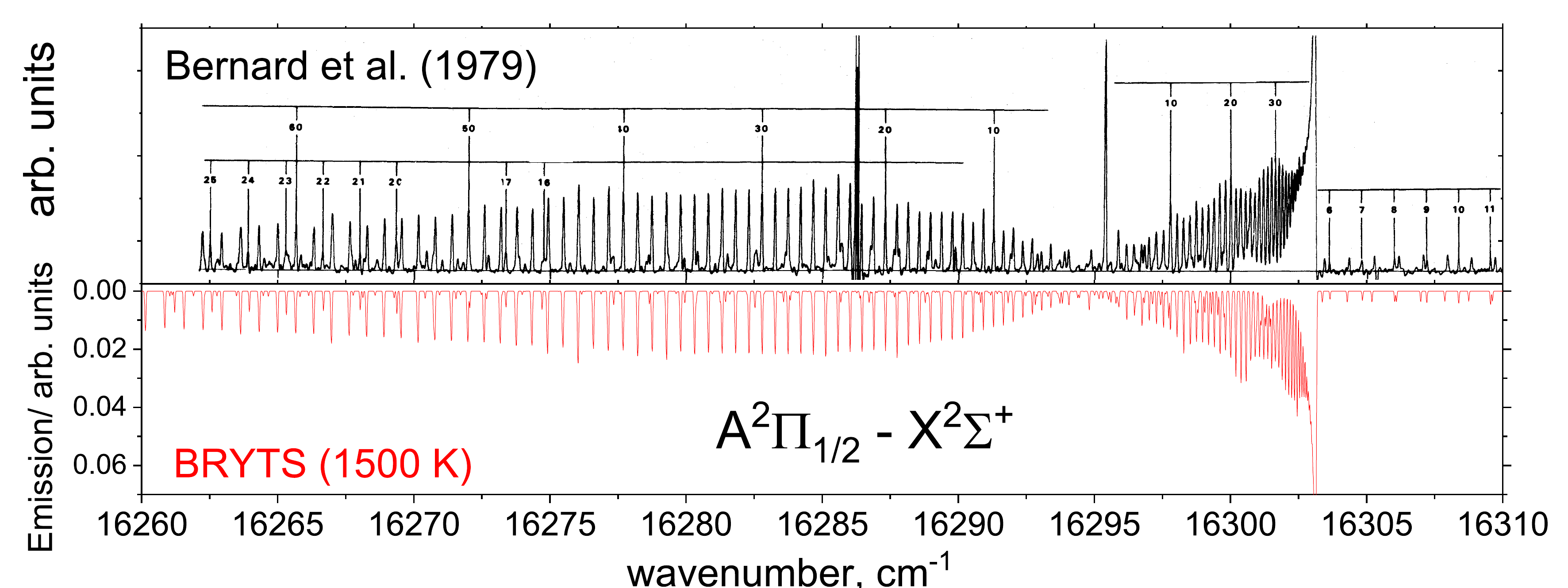}
    \includegraphics[width=\linewidth]{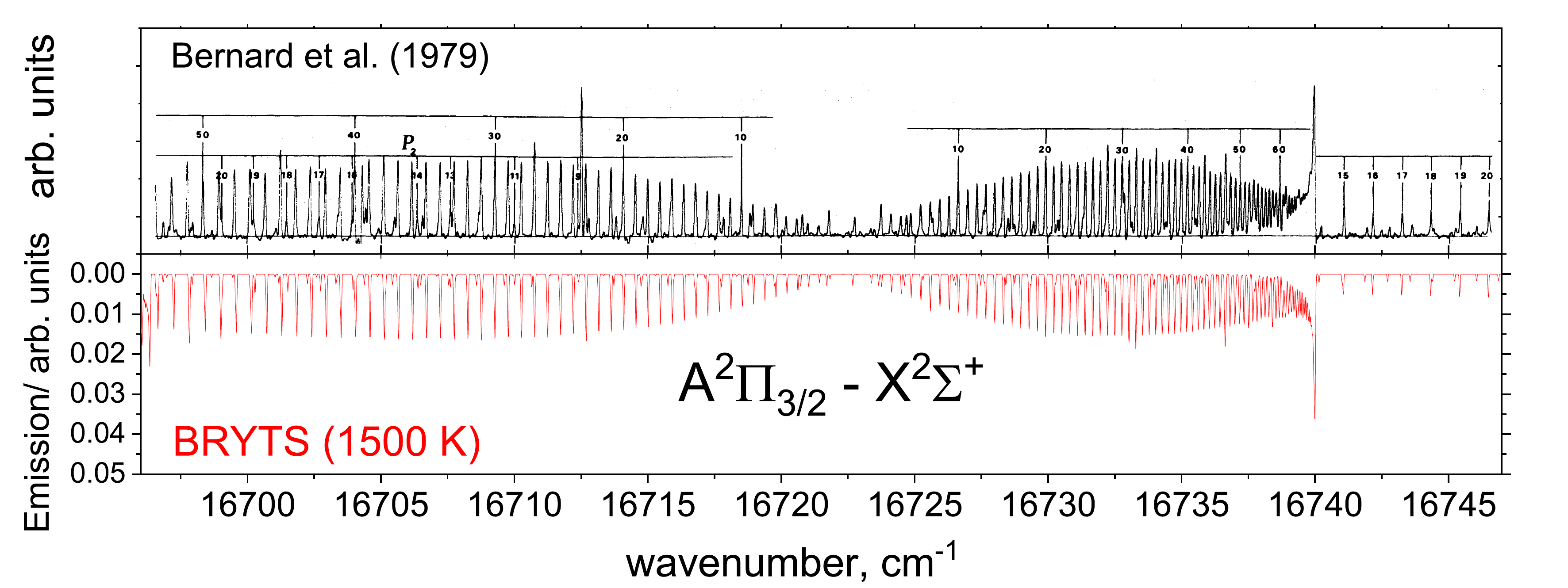}

    \caption{Comparison of our computed \AS$_{1/2}$ $\rightarrow$ \XS\  and \AS$_{3/2}$ $\rightarrow$ \XS\  rotational-electronic emission bands (79BeBaLu, red lines) with those of \citet{79BeBaLu.YO}, a measured spectrum using Fourier Transform Spectroscopy (upper displays). We compute our spectrum at a temperature of 1500~K to best match that of \citet{79BeBaLu.YO}. A Gaussian line profile of 0.065~\cm\ was used with a resolution of 0.01~\cm.   \copyright  AAS. Reproduced with permission.
    }
    \label{fig:79BeBaLu_Comparison}
\end{figure*}

Figure \ref{fig:79BeBaLu_Comparison} compares the experimental  \band{\AS$_{1/2}$}{\XS} $v=0\rightarrow 0$ emission bands measured by \citet{79BeBaLu.YO} via Fourier Transform spectroscopy (black, extracted  from their Fig.~2) to our computed spectra (red). We simulate our spectra at the  temperature of 1500~K to agree with the rotational structure of the experiment.  We see excellent agreement in both line position and band structure with the experiment.
Some discrepancies can be seen in the line intensities, but this could be due to our assumptions about the temperature and line broadening.
Table~\ref{t:bandheads} provides an extended comparisons of the positions of band heads of the \AS--\XS\ system between the experimental values reported by \cite{83BeGrxx.YO} and the theoretical values from the present work, as well as the corresponding values of  $J_{\rm head}$.  Generally, the magnitude of residuals $\Delta \tilde\nu = \tilde{\nu}_{\rm exp}-\tilde{\nu}_{\rm calc}$  correlates with the level of the rotational and vibrational excitations. The band heads with  $J_{\rm head}$ below $70.5$ and $v\le 8$ agree within $\sim 0.1$~\cm, which then degrades to about 0.5~\cm\ for $J_{\rm head} \to 100.5$.

\begin{figure}
	\includegraphics[width=0.5\textwidth]{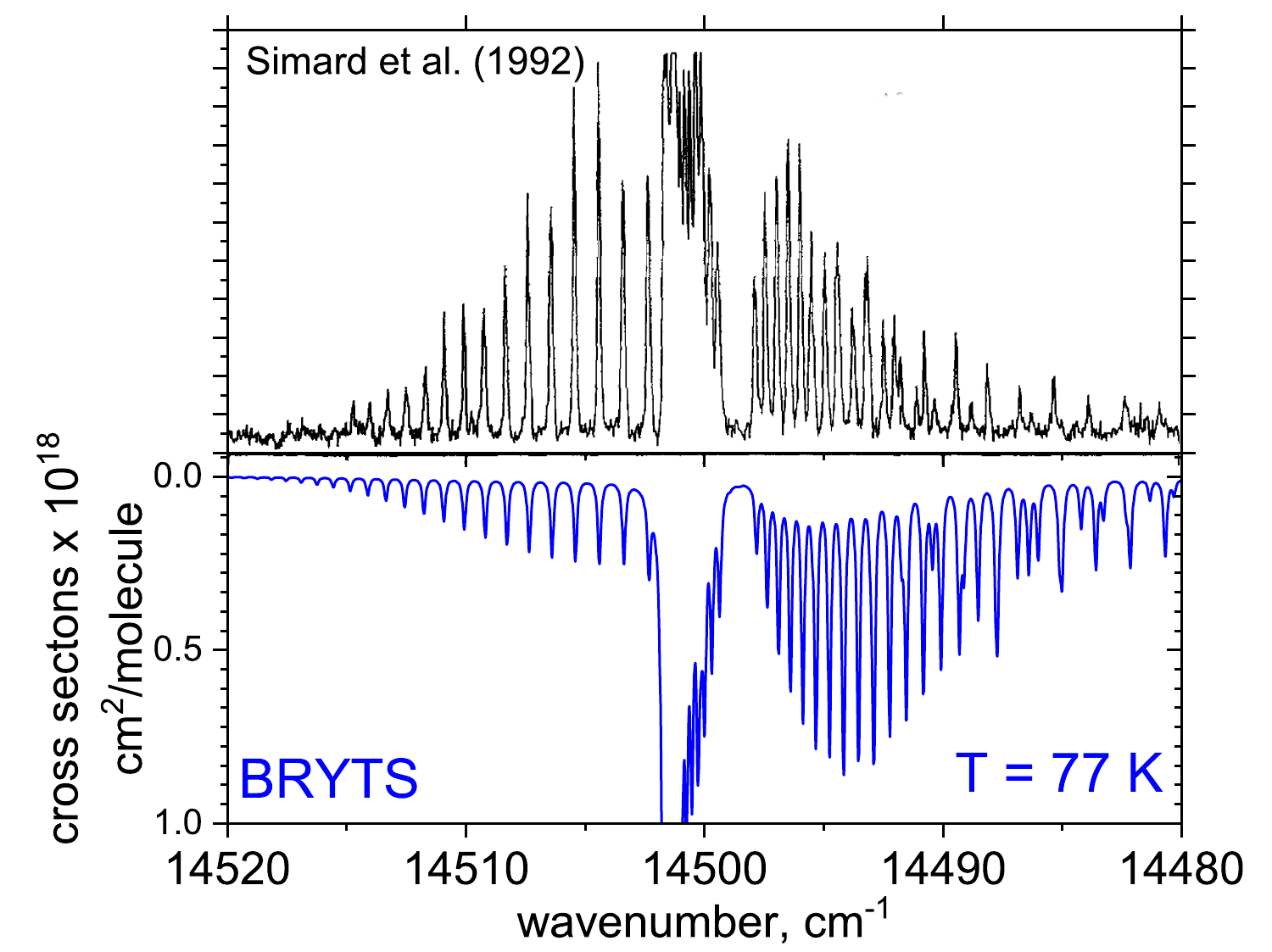}
	\caption{Comparison of the computed emission \ApS\ -- \XS\ (0,0) band with the measurements of
\protect\citet{92SiJaHa.YO} at $T=77$ K and Lorentzian line profile of HWHM = 0.04~\cm.}
	\label{f:T=77K:Ap}
\end{figure}
\begin{figure}
	\includegraphics[width=0.5\textwidth]{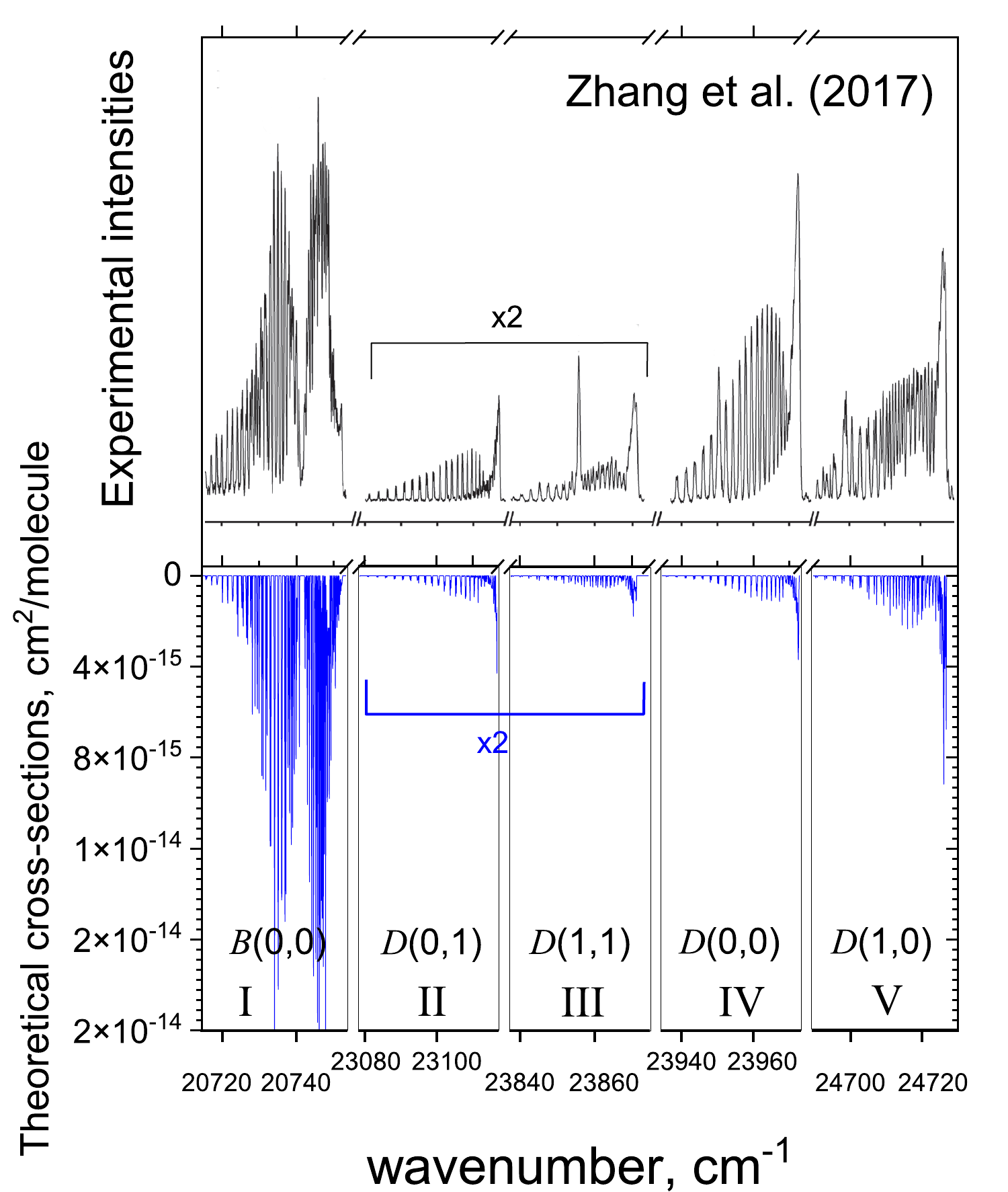}
	\caption{Comparison of our computed (bottom) \DS--\XS\ absorption spectra to the measurements of \citet{17ZhZhZh.YO} (top). The  simulations assumed a cold rotational temperature of $T_{\rm rot}$ = 50 K and a hot vibrational temperature of $T_{\rm vib}$  = 800~K. A Doppler line profile corresponding to $T_{\rm rot}$ = 50 K was used. }
	\label{f:T=77K:2017}
\end{figure}

As in \citet{19SmSoYu}, for the sake of completeness, we provide comparisons with the emission \ApS\ -- \XS\ spectrum from \citet{92SiJaHa.YO}, Fig.~\ref{f:T=77K:Ap}, and the \BS\ -- \XS\ and \DS\ -- \XS\ absorption spectra from \citet{17ZhZhZh.YO},  Fig.~\ref{f:T=77K:2017}, using the \name\ line list, now with an improved agreement, see also the corresponding discussions in \citet{19SmSoYu}. Figure   Fig.~\ref{f:T=77K:2017} shows the simulation of the the forbidden band \ApS\ -- \XS\ (0,0) in emission. In Fig.~\ref{f:T=77K:2017},  non-local thermodynamic equilibrium conditions   with the rotational temperature $T_{\rm rot}$ = 50~K  and the vibrational temperature of $T_{\rm vib}$  = 800~K to better reproduce the experimental spectrum were assumed.

\begin{table*}
\centering
\caption{Comparison of the experimental \citep{83BeGrxx.YO} and theoretical (this work) positions of the band heads in the \AS--\XS\ system. }
\label{t:bandheads}
\begin{tabular}{rrcccrrr} \hline \hline
$v'$  & $v''$& branch  &  $J_{\rm head}$ (obs)  & $J_{\rm head}$ (calc) & $\tilde{\nu}_{\rm exp}$ & $\tilde{\nu}_{\rm calc}$     & $ \Delta \tilde{\nu}$   \\
\hline
   0&   1& $Q_1$ &  93.5  & 98.5 &    15456.10&     15456.61&        -0.51   \\
   1&   2& $Q_1$ &  87.5  & 88.5 &    15417.24&     15417.45&        -0.21   \\
   2&   3& $Q_1$ &  77.5  & 79.5 &    15377.56&     15377.77&        -0.21   \\
   3&   4& $Q_1$ &  70.5  & 71.5 &    15337.30&     15337.45&        -0.15   \\
   4&   5& $Q_1$ &  62.5  & 63.5 &    15296.26&     15296.37&        -0.11   \\
   5&   6& $Q_1$ &  56.5  & 56.5 &    15254.24&     15254.41&        -0.17   \\
   6&   7& $Q_1$ &  49.5  & 50.5 &    15210.78&     15211.28&        -0.50   \\
   8&   9& $Q_1$ &  37.5  & 39.5 &    15123.48&     15123.34&         0.14   \\
   9&  10& $Q_1$ &  33.5  & 35.5 &    15077.15&     15077.28&        -0.13   \\
  10&  11& $Q_1$ &        & 27.5 &    15033.33&     15030.30&         3.03   \\
  11&  12& $Q_1$ &        & 25.5 &    14981.00&     14981.17&        -0.17   \\
  12&  13& $Q_1$ &        & 26.5 &    14933.83&     14934.60&        -0.77   \\
  13&  14& $Q_1$ &        & 23.5 &    14880.28&     14882.94&        -2.66   \\
  14&  15& $Q_1$ &        & 19.5 &    14829.85&     14832.44&        -2.59   \\
  15&  16& $Q_1$ &        & 18.5 &    14774.01&     14778.50&        -4.49   \\
  16&  17& $Q_1$ &        & 18.5 &    14718.59&     14725.27&        -6.68   \\
  17&  18& $Q_1$ &        & 17.5 &    14654.49&     14671.33&       -16.84   \\
  18&  19& $Q_1$ &        & 16.5 &    14615.68&     14616.53&        -0.85   \\
  20&  21& $Q_1$ &        & 63.5 &    14535.74&     14541.80&        -6.06   \\
  21&  22& $Q_1$ &        & 59.5 &    14494.52&     14479.62&        14.90   \\
  22&  23& $Q_1$ &        & 54.5 &    14452.30&     14416.66&        35.64   \\
   0&   0& $Q_1$ &  51.5  & 51.5 &    16303.10&     16303.16&        -0.06   \\
   0&   0& $R_2$ &  79.5  & 81.5 &    16739.90&     16740.01&        -0.11   \\
   1&   1& $Q_1$ &  47.5  & 48.5 &    16259.93&     16259.89&         0.04   \\
   1&   1& $R_2$ &  72.5  & 74.5 &    16696.30&     16696.36&        -0.06   \\
   2&   2& $Q_1$ &  44.5  & 44.5 &    16215.91&     16215.89&         0.02   \\
   2&   2& $R_2$ &  66.5  & 68.5 &    16652.15&     16652.10&         0.05   \\
   3&   3& $Q_1$ &  40.5  & 41.5 &    16171.00&     16171.02&        -0.02   \\
   3&   3& $R_2$ &  61.5  & 62.5 &    16607.07&     16607.07&         0.00   \\
   4&   4& $Q_1$ &  37.5  & 37.5 &    16125.08&     16125.20&        -0.12   \\
   4&   4& $R_2$ &  56.5  & 57.5 &    16561.11&     16561.16&        -0.05   \\
   5&   5& $Q_1$ &  34.5  & 35.5 &    16078.10&     16078.30&        -0.20   \\
   5&   5& $R_2$ &  51.5  & 52.5 &    16514.06&     16514.08&        -0.02   \\
   6&   6& $Q_1$ &  31.5  & 32.5 &    16029.60&     16030.15&        -0.55   \\
   6&   6& $R_2$ &        & 35.5 &    16462.40&     16462.74&        -0.34   \\
   7&   7& $R_2$ &  43.5  & 44.5 &    16417.60&     16418.85&        -1.25   \\
   8&   8& $Q_1$ &  26.5  & 27.5 &    15932.00&     15931.69&         0.31   \\
   8&   8& $R_2$ &  37.5  & 37.5 &    16367.89&     16370.33&        -2.44   \\
   9&   9& $Q_1$ &  24.5  & 25.5 &    15880.20&     15880.17&         0.03   \\
   9&   9& $R_2$ &  33.5  & 35.5 &    16315.88&     16317.95&        -2.07   \\
  10&  10& $Q_1$ &        & 22.5 &    15828.06&     15827.76&         0.30   \\
  11&  11& $Q_1$ &        & 19.5 &    15773.23&     15773.03&         0.20   \\
  12&  12& $Q_1$ &        & 20.5 &    15720.30&     15720.47&        -0.17   \\
  \hline
   \hline
\end{tabular}
\end{table*}

\section{Conclusions}

Accurate and extensive empirical  \name\ line lists for $^{89}$Y$^{16}$O, $^{89}$Y$^{17}$O  and $^{89}$Y$^{18}$O  are produced covering six lowest doublet electronic states and ranging up to 60~000~\cm. The line list is based on a refined set of curves in the diabatic representation obtained by fitting to a set of experimentally derived rovibronic energies of YO.  The latter is  based  on  the experimental data from the literature, either  original laboratory line positions whenever available or spectroscopic constants. Using an effective Hamiltonian to reconstruct molecular energies in place of the original experimental data is less than ideal as it lacks information on any local perturbations, which is critical when using it to fit the spectroscopic model.

Although  ExoMol line lists, including \name, are usually intended for astrophysical applications of hot atmospheric environments, YO is one of the molecules used in cooling applications, where our line list may also be useful.

The \ai\ calculations, especially MRCI, of transition metal species are still a big challenge and therefore ultimately   the lab data (transition frequencies, intensities, dipoles, lifetimes) is a crucial source of the information to produce useful line lists. For YO we were lucky to have the \ai\ PECs  of excited electronic states of the CCSD(T) quality, while everything else had to rely on the fit to the experiment.

In this work, the hyperfine structure of the YO rovibronic states was ignored, mostly due to the lack of the experiment. Should it become important for YO spectroscopic applications  to include the hyperfine effects, the methodology to compute the hyperfine-resolved energies and spectra is readily available as implemented in \Duo\ \citep{jt855,jt873,jt912}.

\section*{Acknowledgements}

We thank Amanda Ross for extremely valuable  advice on the effective rotational Hamiltonian in connection with the  \AS\--\XS\ data by  Bernard et al. (1979). Her help led to huge improvement of our \AS\ state model and in the associated quality of the line list. This work was  supported by the European Research Council
(ERC) under the European Union’s Horizon 2020 research and innovation programme through Advance Grant number 883830 and  the STFC Projects No. ST/M001334/1 and ST/R000476/1. The authors acknowledge the use of the Cambridge Service for Data Driven Discovery (CSD3) as part of the STFC DiRAC HPC Facility (www.dirac.ac.uk),  funded by BEIS capital funding via STFC capital grants ST/P002307/1 and ST/R002452/1 and STFC operations grant ST/R00689X/1. A.N.S. and V.G.S. acknowledge support from Project No. FZZW-2023-0010.

\section*{Data Availability}

The states, transition, opacity and partition function files for the YO line lists can be downloaded from \href{https://exomol.com}{www.exomol.com}. The open access programs \Duo\ and \textsc{ExoCross} are available from \href{https://github.com/exomol}{github.com/exomol}.

\section*{Supporting Information}

Supplementary data are available at MNRAS online. This includes (i) the spectroscopic model in the form of the \Duo\ input file, containing all the curves, parameters as well as the experimentally derived energy term values of YO used in the fit; (ii) the experimental line positions collected from the literature in the MARVEL format and (iii) an effective Hamiltonian for a $^2\Pi$ electronic state from \citet{77BaCeDI.CN}.




\begin{thebibliography}{}
\makeatletter
\relax
\def\mn@urlcharsother{\let\do\@makeother \do\$\do\&\do\#\do\^\do\_\do\%\do\~}
\def\mn@doi{\begingroup\mn@urlcharsother \@ifnextchar [ {\mn@doi@}
  {\mn@doi@[]}}
\def\mn@doi@[#1]#2{\def\@tempa{#1}\ifx\@tempa\@empty \href
  {http://dx.doi.org/#2} {doi:#2}\else \href {http://dx.doi.org/#2} {#1}\fi
  \endgroup}
\def\mn@eprint#1#2{\mn@eprint@#1:#2::\@nil}
\def\mn@eprint@arXiv#1{\href {http://arxiv.org/abs/#1} {{\tt arXiv:#1}}}
\def\mn@eprint@dblp#1{\href {http://dblp.uni-trier.de/rec/bibtex/#1.xml}
  {dblp:#1}}
\def\mn@eprint@#1:#2:#3:#4\@nil{\def\@tempa {#1}\def\@tempb {#2}\def\@tempc
  {#3}\ifx \@tempc \@empty \let \@tempc \@tempb \let \@tempb \@tempa \fi \ifx
  \@tempb \@empty \def\@tempb {arXiv}\fi \@ifundefined
  {mn@eprint@\@tempb}{\@tempb:\@tempc}{\expandafter \expandafter \csname
  mn@eprint@\@tempb\endcsname \expandafter{\@tempc}}}

\bibitem[\protect\citeauthoryear{Ackermann \& Rauh}{Ackermann \&
  Rauh}{1974}]{74AcRa}
Ackermann R.~J.,  Rauh E.~G.,  1974, \mn@doi [J. Chem. Phys.]
  {10.1063/1.1681357}, 60, 2266

\bibitem[\protect\citeauthoryear{Al-Refaie, Changeat, Waldmann  \&
  Tinetti}{Al-Refaie et~al.}{2021}]{TauRex3}
Al-Refaie A.~F.,  Changeat Q.,  Waldmann I.~P.,   Tinetti G.,  2021, \mn@doi
  [ApJ] {10.3847/1538-4357/ac0252}, 917, 37

\bibitem[\protect\citeauthoryear{Bacis, Cerny, D'Incan, Guelachvili  \&
  Roux}{Bacis et~al.}{1977}]{77BaCeDI.CN}
Bacis R.,  Cerny D.,  D'Incan J.,  Guelachvili G.,   Roux F.,  1977, \mn@doi
  [ApJ] {10.1086/155327}, 214, 946

\bibitem[\protect\citeauthoryear{Badie \& Granier}{Badie \&
  Granier}{2002}]{02BaGrxx.YO}
Badie J.~M.,  Granier B.,  2002, \mn@doi [Chem. Phys. Lett.]
  {10.1016/S0009-2614(02)01366-0}, 364, 550

\bibitem[\protect\citeauthoryear{Badie \& Granier}{Badie \&
  Granier}{2003}]{03BaGrxx.YO}
Badie J.~M.,  Granier B.,  2003, \mn@doi [Eur. Phys. J.-Appl. Phys]
  {10.1051/epjap:2003007}, 21, 239

\bibitem[\protect\citeauthoryear{Badie, Cassan  \& Granier}{Badie
  et~al.}{2005a}]{05BaCaG1.YO}
Badie J.~M.,  Cassan L.,   Granier B.,  2005a, \mn@doi [Eur. Phys. J.-Appl.
  Phys] {10.1051/epjap:2004202}, 29, 111

\bibitem[\protect\citeauthoryear{Badie, Cassan  \& Granier}{Badie
  et~al.}{2005b}]{05BaCaGr.YO}
Badie J.~M.,  Cassan L.,   Granier B.,  2005b, \mn@doi [Eur. Phys. J.-Appl.
  Phys] {10.1051/epjap:2005070}, 32, 61

\bibitem[\protect\citeauthoryear{Badie, Cassan  \& Granier}{Badie
  et~al.}{2007a}]{07BaCaG1.YO}
Badie J.~M.,  Cassan L.,   Granier B.,  2007a, \mn@doi [Eur. Phys. J.-Appl.
  Phys] {10.1051/epjap:2007059}, 38, 177

\bibitem[\protect\citeauthoryear{Badie, Cassan, Granier, Florez  \&
  Janna}{Badie et~al.}{2007b}]{07BaCaGr.YO}
Badie J.~M.,  Cassan L.,  Granier B.,  Florez S.~A.,   Janna F.~C.,  2007b,
  \mn@doi [J. Sol. Energy Eng. Trans.-ASME] {10.1115/1.2769718}, 129, 412

\bibitem[\protect\citeauthoryear{Bagare \& Murthy}{Bagare \&
  Murthy}{1982}]{82BaMuxx.YO}
Bagare S.~P.,  Murthy N.~S.,  1982, \mn@doi [Pramana] {10.1007/BF02847382}, 19,
  497

\bibitem[\protect\citeauthoryear{Barklem \& Collet}{Barklem \&
  Collet}{2016}]{16BaCoxx.partfunc}
Barklem P.~S.,  Collet R.,  2016, \mn@doi [A\&A] {10.1051/0004-6361/201526961},
  588, A96

\bibitem[\protect\citeauthoryear{Bernard \& Gravina}{Bernard \&
  Gravina}{1980}]{80BeGrxx.YO}
Bernard A.,  Gravina R.,  1980, \mn@doi [ApJS] {10.1086/190692}, 44, 223

\bibitem[\protect\citeauthoryear{Bernard \& Gravina}{Bernard \&
  Gravina}{1983}]{83BeGrxx.YO}
Bernard A.,  Gravina R.,  1983, \mn@doi [ApJS] {10.1086/190877}, 52, 443

\bibitem[\protect\citeauthoryear{Bernard, Bacis  \& Luc}{Bernard
  et~al.}{1979}]{79BeBaLu.YO}
Bernard A.,  Bacis R.,   Luc P.,  1979, \mn@doi [ApJ] {10.1086/156736}, 227,
  338

\bibitem[\protect\citeauthoryear{Bowesman, Yurchenko  \& Tennyson}{Bowesman
  et~al.}{2023}]{jt912}
Bowesman C.~A.,  Yurchenko S.~N.,   Tennyson J.,  2023, \mn@doi [Mol. Phys.]
  {10.1080/00268976.2023.2255299}

\bibitem[\protect\citeauthoryear{Brady, Yurchenko, Kim, Somogyi  \&
  Tennyson}{Brady et~al.}{2022}]{22BrYuKi}
Brady R.~P.,  Yurchenko S.~N.,  Kim G.-S.,  Somogyi W.,   Tennyson J.,  2022,
  \mn@doi [Phys. Chem. Chem. Phys.] {10.1039/D2CP03051A}, 24, 24076

\bibitem[\protect\citeauthoryear{Brady, Drury, Tennyson  \& Yurchenko}{Brady
  et~al.}{2023}]{23BrDrTe}
Brady R.~P.,  Drury C.,  Tennyson J.,   Yurchenko S.~N.,  2023, Mol. Phys., in
  preparation

\bibitem[\protect\citeauthoryear{Brown \& Merer}{Brown \&
  Merer}{1979}]{79BrMexx.methods}
Brown J.~M.,  Merer A.~J.,  1979, \mn@doi [J. Mol. Spectrosc.]
  {10.1016/0022-2852(79)90172-3}, 74, 488

\bibitem[\protect\citeauthoryear{Cazzoli, Cludi  \& Puzzarini}{Cazzoli
  et~al.}{2006}]{06CaClLi.PN}
Cazzoli G.,  Cludi L.,   Puzzarini C.,  2006, \mn@doi [J. Mol. Struct.]
  {10.1016/j.molstruc.2005.07.010}, 780-81, 260

\bibitem[\protect\citeauthoryear{Chalek \& Gole}{Chalek \&
  Gole}{1976}]{76ChGoxx.YO}
Chalek C.~L.,  Gole J.~L.,  1976, \mn@doi [J. Chem. Phys.] {10.1063/1.433434},
  65, 2845

\bibitem[\protect\citeauthoryear{Chalek \& Gole}{Chalek \&
  Gole}{1977}]{77ChGoxx.YO}
Chalek C.~L.,  Gole J.~L.,  1977, \mn@doi [Chem. Phys.]
  {10.1016/0301-0104(77)80007-4}, 19, 59

\bibitem[\protect\citeauthoryear{Childs, Poulsen  \& Steimle}{Childs
  et~al.}{1988}]{88ChPoSt.YO}
Childs W.~J.,  Poulsen O.,   Steimle T.~C.,  1988, \mn@doi [J. Chem. Phys.]
  {10.1063/1.454186}, 88, 598

\bibitem[\protect\citeauthoryear{Chubb et~al.,}{Chubb et~al.}{2021}]{jt801}
Chubb K.~L.,  et~al., 2021, \mn@doi [A\&A] {10.1051/0004-6361/202038350}, 646,
  A21

\bibitem[\protect\citeauthoryear{Collopy, Hummon, Yeo, Yan  \& Ye}{Collopy
  et~al.}{2015}]{15CoHuYe.YO}
Collopy A.~L.,  Hummon M.~T.,  Yeo M.,  Yan B.,   Ye J.,  2015, \mn@doi [New J.
  Phys] {10.1088/1367-2630/17/5/055008}, 17, 055008

\bibitem[\protect\citeauthoryear{Collopy, Ding, Wu, Finneran, Anderegg,
  Augenbraun, Doyle  \& Ye}{Collopy et~al.}{2018}]{18CoDiWu.YO}
Collopy A.~L.,  Ding S.,  Wu Y.,  Finneran I.~A.,  Anderegg L.,  Augenbraun
  B.~L.,  Doyle J.~M.,   Ye J.,  2018, \mn@doi [Phys. Rev. Lett.]
  {10.1103/PhysRevLett.121.213201}, 121, 213201

\bibitem[\protect\citeauthoryear{Dye, Muenchausen  \& Nogar}{Dye
  et~al.}{1991}]{91DyMuNo.YO}
Dye R.~C.,  Muenchausen R.~E.,   Nogar N.~S.,  1991, \mn@doi [Chem. Phys.
  Lett.] {10.1016/0009-2614(91)80308-K}, 181, 531

\bibitem[\protect\citeauthoryear{Fried, Kushida, Reck  \& Rothe}{Fried
  et~al.}{1993}]{93FrKuRe.YO}
Fried D.,  Kushida T.,  Reck G.~P.,   Rothe E.~W.,  1993, \mn@doi [J. Appl.
  Phys.] {10.1063/1.353955}, 73, 7810

\bibitem[\protect\citeauthoryear{Furtenbacher, {Cs\'asz\'ar}  \&
  Tennyson}{Furtenbacher et~al.}{2007}]{MARVEL}
Furtenbacher T.,  {Cs\'asz\'ar} A.~G.,   Tennyson J.,  2007, \mn@doi [J. Mol.
  Spectrosc.] {10.1016/j.jms.2007.07.005}, 245, 115

\bibitem[\protect\citeauthoryear{Goranskii \& Barsukova}{Goranskii \&
  Barsukova}{2007}]{07GoBaxx.YO}
Goranskii V.~P.,  Barsukova E.~A.,  2007, \mn@doi [Astron. Rep.]
  {10.1134/S1063772907020072}, 51, 126

\bibitem[\protect\citeauthoryear{Hoeft \& Torring}{Hoeft \&
  Torring}{1993}]{93HoToxx.YO}
Hoeft J.,  Torring T.,  1993, \mn@doi [Chem. Phys. Lett.]
  {10.1016/0009-2614(93)85730-C}, 215, 367

\bibitem[\protect\citeauthoryear{Hulburt \& Hirschfelder}{Hulburt \&
  Hirschfelder}{1941}]{41HuHi}
Hulburt H.~M.,  Hirschfelder J.~O.,  1941, \mn@doi [J. Chem. Phys.]
  {10.1063/1.1750827}, 9, 61

\bibitem[\protect\citeauthoryear{Irwin et~al.,}{Irwin et~al.}{2008}]{NEMESIS}
Irwin P. G.~J.,  et~al., 2008, \mn@doi [J. Quant. Spectrosc. Radiat. Transf.]
  {10.1016/j.jqsrt.2007.11.006}, 109, 1136

\bibitem[\protect\citeauthoryear{Kaminski, Schmidt, Tylenda, Konacki  \&
  Gromadzki}{Kaminski et~al.}{2009}]{09KaScTy.YO}
Kaminski T.,  Schmidt M.,  Tylenda R.,  Konacki M.,   Gromadzki M.,  2009,
  \mn@doi [ApJS] {10.1088/0067-0049/182/1/33}, 182, 33

\bibitem[\protect\citeauthoryear{Kasai \& Weltner}{Kasai \&
  Weltner}{1965}]{65KaWexx.YO}
Kasai P.~H.,  Weltner Jr. W.,  1965, \mn@doi [J. Chem. Phys.]
  {10.1063/1.1697161}, 43, 2553

\bibitem[\protect\citeauthoryear{Knight, Kaup, Petzoldt, Ayyad, Ghanty  \&
  Davidson}{Knight et~al.}{1999}]{99KnKaPe.YO}
Knight L.~B.,  Kaup J.~G.,  Petzoldt B.,  Ayyad R.,  Ghanty T.~K.,   Davidson
  E.~R.,  1999, \mn@doi [J. Chem. Phys.] {10.1063/1.478464}, 110, 5658

\bibitem[\protect\citeauthoryear{Kobayashi \& Sekine}{Kobayashi \&
  Sekine}{2006}]{06KoSexx.YO}
Kobayashi T.,  Sekine T.,  2006, \mn@doi [Chem. Phys. Lett.]
  {10.1016/j.cplett.2006.04.056}, 424, 54

\bibitem[\protect\citeauthoryear{Langhoff \& Bauschlicher}{Langhoff \&
  Bauschlicher}{1988}]{88LaBaxx.YO}
Langhoff S.~R.,  Bauschlicher C.~W.,  1988, \mn@doi [J. Chem. Phys.]
  {10.1063/1.455060}, 89, 2160

\bibitem[\protect\citeauthoryear{Lee, Seto, Hirao, Bernath  \& Le~Roy}{Lee
  et~al.}{1999}]{EMO}
Lee E.~G.,  Seto J.~Y.,  Hirao T.,  Bernath P.~F.,   Le~Roy R.~J.,  1999,
  \mn@doi [J. Mol. Spectrosc.] {10.1006/jmsp.1998.7789}, 194, 197

\bibitem[\protect\citeauthoryear{Leung, Ma  \& Cheung}{Leung
  et~al.}{2005}]{05LeMaCh.YO}
Leung J. W.~H.,  Ma T.~M.,   Cheung A. S.~C.,  2005, \mn@doi [J. Mol.
  Spectrosc.] {10.1016/j.jms.2004.08.017}, 229, 108

\bibitem[\protect\citeauthoryear{Linton}{Linton}{1978}]{78Linton.YO}
Linton C.,  1978, \mn@doi [J. Mol. Spectrosc.] {10.1016/0022-2852(78)90228-X},
  69, 351

\bibitem[\protect\citeauthoryear{Liu \& Parson}{Liu \&
  Parson}{1977}]{77LiPaxx.YO}
Liu K.,  Parson J.~M.,  1977, \mn@doi [J. Chem. Phys.] {10.1063/1.435138}, 67,
  1814

\bibitem[\protect\citeauthoryear{Liu \& Parson}{Liu \&
  Parson}{1979}]{79LiPaxx.YO}
Liu K.,  Parson J.~M.,  1979, \mn@doi [J. Phys. Chem.] {10.1021/j100471a017},
  83, 970

\bibitem[\protect\citeauthoryear{Manos \& Parson}{Manos \&
  Parson}{1975}]{75MaPaxx.YO}
Manos D.~M.,  Parson J.~M.,  1975, \mn@doi [J. Chem. Phys.] {10.1063/1.431798},
  63, 3575

\bibitem[\protect\citeauthoryear{Medvedev}{Medvedev}{2012}]{12Medvedev}
Medvedev E.~S.,  2012, \mn@doi [J. Chem. Phys.] {10.1063/1.4761930}, 137,
  174307

\bibitem[\protect\citeauthoryear{Medvedev \& Ushakov}{Medvedev \&
  Ushakov}{2022}]{22MeUs.CO}
Medvedev E.~S.,  Ushakov V.~G.,  2022, \mn@doi [J. Quant. Spectrosc. Radiat.
  Transf.] {10.1016/j.jqsrt.2022.108255}, 288, 108255

\bibitem[\protect\citeauthoryear{Medvedev, Meshkov, Stolyarov  \&
  Gordon}{Medvedev et~al.}{2015}]{15MeMeSt.CO}
Medvedev E.~S.,  Meshkov V.~V.,  Stolyarov A.~V.,   Gordon I.~E.,  2015,
  \mn@doi [J. Chem. Phys.] {10.1063/1.4933136}, 143, 154301

\bibitem[\protect\citeauthoryear{Medvedev, Meshkov, Stolyarov, Ushakov  \&
  Gordon}{Medvedev et~al.}{2016}]{16MeMeSt}
Medvedev E.~S.,  Meshkov V.~V.,  Stolyarov A.~V.,  Ushakov V.~G.,   Gordon
  I.~E.,  2016, \mn@doi [J. Mol. Spectrosc.] {10.1016/j.jms.2016.06.013}, 330,
  36

\bibitem[\protect\citeauthoryear{Min, Ormel, Chubb, Helling  \& Kawashima}{Min
  et~al.}{2020}]{ARCiS}
Min M.,  Ormel C.~W.,  Chubb K.,  Helling C.,   Kawashima Y.,  2020, \mn@doi
  [A\&A] {10.1051/0004-6361/201937377}, 642, A28

\bibitem[\protect\citeauthoryear{Molli{\'e}re, Wardenier, {van Boekel},
  Henning, Molaverdikhani  \& Snellen}{Molli{\'e}re
  et~al.}{2019}]{19MoWaBo.petitRADTRANS}
Molli{\'e}re P.,  Wardenier J.~P.,  {van Boekel} R.,  Henning T.,
  Molaverdikhani K.,   Snellen I. A.~G.,  2019, \mn@doi [A\&A]
  {10.1051/0004-6361/201935470}, 627, A67

\bibitem[\protect\citeauthoryear{Mukund \& Nakhate}{Mukund \&
  Nakhate}{2023}]{23MuNa.YO}
Mukund S.,  Nakhate S.~G.,  2023, \mn@doi [J. Quant. Spectrosc. Radiat.
  Transf.] {10.1016/j.jqsrt.2022.108452}, 296, 108452

\bibitem[\protect\citeauthoryear{Murty}{Murty}{1982}]{82Murty.YO}
Murty P.~S.,  1982, Astrophys. Lett., 23, 7

\bibitem[\protect\citeauthoryear{Murty}{Murty}{1983}]{83Murty.YO}
Murty P.~S.,  1983, \mn@doi [Astrophys. Space Sci.] {10.1007/BF00653719}, 94,
  295

\bibitem[\protect\citeauthoryear{Otis \& Goodwin}{Otis \&
  Goodwin}{1993}]{93OtGoxx.YO}
Otis C.~E.,  Goodwin P.~M.,  1993, \mn@doi [J. Appl. Phys.] {10.1063/1.353186},
  73, 1957

\bibitem[\protect\citeauthoryear{Peterson \& Dunning}{Peterson \&
  Dunning}{2002}]{02PeDuxx.ai}
Peterson K.~A.,  Dunning T.~H.,  2002, \mn@doi [J. Chem. Phys.]
  {10.1063/1.1520138}, 117, 10548

\bibitem[\protect\citeauthoryear{Peterson, Figgen, Dolg  \& Stoll}{Peterson
  et~al.}{2007}]{07PeFiDo.ai}
Peterson K.~A.,  Figgen D.,  Dolg M.,   Stoll H.,  2007, \mn@doi [J. Chem.
  Phys.] {10.1063/1.2647019}, 126, 124101

\bibitem[\protect\citeauthoryear{Prajapat, Jagoda, Lodi, Gorman, Yurchenko  \&
  Tennyson}{Prajapat et~al.}{2017}]{jt703}
Prajapat L.,  Jagoda P.,  Lodi L.,  Gorman M.~N.,  Yurchenko S.~N.,   Tennyson
  J.,  2017, \mn@doi [MNRAS] {10.1093/mnras/stx2229}, 472, 3648

\bibitem[\protect\citeauthoryear{Qu, Yurchenko  \& Tennyson}{Qu
  et~al.}{2022a}]{jt855}
Qu Q.,  Yurchenko S.~N.,   Tennyson J.,  2022a, \mn@doi [J. Chem. Theory
  Comput.] {10.1021/acs.jctc.1c01244}, 18, 1808

\bibitem[\protect\citeauthoryear{Qu, Yurchenko  \& Tennyson}{Qu
  et~al.}{2022b}]{jt873}
Qu Q.,  Yurchenko S.~N.,   Tennyson J.,  2022b, \mn@doi [J. Chem. Phys.]
  {10.1063/5.0105965}, 157, 124305

\bibitem[\protect\citeauthoryear{Qu{\'e}m{\'e}ner \& Bohn}{Qu{\'e}m{\'e}ner \&
  Bohn}{2016}]{16QuGoJo.YO}
Qu{\'e}m{\'e}ner G.,  Bohn J.~L.,  2016, \mn@doi [Phys. Rev. A]
  {10.1103/PhysRevA.93.012704}, 93, 012704

\bibitem[\protect\citeauthoryear{Sauval}{Sauval}{1978}]{78Sauval}
Sauval A.~J.,  1978, A\&A, 62, 295

\bibitem[\protect\citeauthoryear{Semenov, Yurchenko  \& Tennyson}{Semenov
  et~al.}{2016}]{16SeYuTe}
Semenov M.,  Yurchenko S.~N.,   Tennyson J.,  2016, \mn@doi [J. Mol.
  Spectrosc.] {10.1016/j.jms.2016.11.004}, 330, 57

\bibitem[\protect\citeauthoryear{Shin \& Nicholls}{Shin \&
  Nicholls}{1977}]{77ShNixx.YO}
Shin J.~B.,  Nicholls R.~W.,  1977, \mn@doi [Spectr. Lett.]
  {10.1080/00387017708065029}, 10, 923

\bibitem[\protect\citeauthoryear{Simard, James, Hackett  \& Balfour}{Simard
  et~al.}{1992}]{92SiJaHa.YO}
Simard B.,  James A.~M.,  Hackett P.~A.,   Balfour W.~J.,  1992, \mn@doi [J.
  Mol. Spectrosc.] {10.1016/0022-2852(92)90225-D}, 154, 455

\bibitem[\protect\citeauthoryear{Smirnov, Solomonik, Yurchenko  \&
  Tennyson}{Smirnov et~al.}{2019}]{19SmSoYu}
Smirnov A.~N.,  Solomonik V.~G.,  Yurchenko S.~N.,   Tennyson J.,  2019,
  \mn@doi [Phys. Chem. Chem. Phys.] {10.1039/C9CP03208H}, 21, 22794

\bibitem[\protect\citeauthoryear{Steimle \& Alramadin}{Steimle \&
  Alramadin}{1986}]{86StAlxx.YO}
Steimle T.~C.,  Alramadin Y.,  1986, \mn@doi [Chem. Phys. Lett.]
  {10.1016/0009-2614(86)80429-8}, 130, 76

\bibitem[\protect\citeauthoryear{Steimle \& Alramadin}{Steimle \&
  Alramadin}{1987}]{87StAlxx.YO}
Steimle T.~C.,  Alramadin Y.,  1987, \mn@doi [J. Mol. Spectrosc.]
  {10.1016/0022-2852(87)90221-9}, 122, 103

\bibitem[\protect\citeauthoryear{Steimle \& Shirley}{Steimle \&
  Shirley}{1990}]{90StShxx.YO}
Steimle T.~C.,  Shirley J.~E.,  1990, \mn@doi [J. Chem. Phys.]
  {10.1063/1.457888}, 92, 3292

\bibitem[\protect\citeauthoryear{Steimle \& Virgo}{Steimle \&
  Virgo}{2003}]{03StVixx.YO}
Steimle T.~C.,  Virgo W.,  2003, \mn@doi [J. Mol. Spectrosc.]
  {10.1016/S0022-2852(03)00165-6}, 221, 57

\bibitem[\protect\citeauthoryear{Suenram, Lovas, Fraser  \& Matsumura}{Suenram
  et~al.}{1990}]{90SuLoFr.YO}
Suenram R.~D.,  Lovas F.~J.,  Fraser G.~T.,   Matsumura K.,  1990, \mn@doi [J.
  Chem. Phys.] {10.1063/1.457690}, 92, 4724

\bibitem[\protect\citeauthoryear{Tennyson \& Yurchenko}{Tennyson \&
  Yurchenko}{2012}]{jt528}
Tennyson J.,  Yurchenko S.~N.,  2012, \mn@doi [MNRAS]
  {10.1111/j.1365-2966.2012.21440.x}, 425, 21

\bibitem[\protect\citeauthoryear{Tennyson, Hill  \& Yurchenko}{Tennyson
  et~al.}{2013}]{jt548}
Tennyson J.,  Hill C.,   Yurchenko S.~N.,  2013, in 6$^{th}$ international
  conference on atomic and molecular data and their applications ICAMDATA-2012.
  AIP, New York, pp 186--195, \mn@doi{10.1063/1.4815853}

\bibitem[\protect\citeauthoryear{Tennyson et~al.,}{Tennyson
  et~al.}{2020}]{jt810}
Tennyson J.,  et~al., 2020, \mn@doi [J. Quant. Spectrosc. Radiat. Transf.]
  {10.1016/j.jqsrt.2020.107228}, 255, 107228

\bibitem[\protect\citeauthoryear{Tylenda, G\'orny, Kam\'{i}nski  \&
  Schmidt}{Tylenda et~al.}{2015}]{15TyGoKa}
Tylenda R.,  G\'orny S.~K.,  Kam\'{i}nski T.,   Schmidt M.,  2015, \mn@doi
  [A\&A] {10.1051/0004-6361/201425592}, 578, A75

\bibitem[\protect\citeauthoryear{Uhler \& Akerlind}{Uhler \&
  Akerlind}{1961}]{61UhAkxx.YO}
Uhler U.,  Akerlind L.,  1961, Arkiv For Fysik, 19, 1

\bibitem[\protect\citeauthoryear{Ushakov, Semenov, Yurchenko, Ermilov  \&
  Medvedev}{Ushakov et~al.}{2023}]{23UsSeYu}
Ushakov V.,  Semenov M.,  Yurchenko S.,  Ermilov A.~Y.,   Medvedev E.,  2023,
  \mn@doi [J. Mol. Spectrosc.] {10.1016/j.jms.2023.111804}, p. 111804

\bibitem[\protect\citeauthoryear{Vardya}{Vardya}{1970}]{70Vardya.YO}
Vardya M.~S.,  1970, A\&A, 5, 162

\bibitem[\protect\citeauthoryear{Werner et~al.,}{Werner
  et~al.}{2020}]{MOLPRO2020}
Werner H.-J.,  et~al., 2020, \mn@doi [The Journal of Chemical Physics]
  {10.1063/5.0005081}, 152, 144107

\bibitem[\protect\citeauthoryear{Western}{Western}{2017}]{PGOPHER}
Western C.~M.,  2017, \mn@doi [J. Quant. Spectrosc. Radiat. Transf.]
  {10.1016/j.jqsrt.2016.04.010}, 186, 221

\bibitem[\protect\citeauthoryear{Wijchers, Dijkerman, Zeegers  \&
  Alkemade}{Wijchers et~al.}{1980}]{80WiDiZe.YO}
Wijchers T.,  Dijkerman H.~A.,  Zeegers P. J.~T.,   Alkemade C. T.~J.,  1980,
  \mn@doi [Spectra Chimica Acta B] {10.1016/0584-8547(80)80090-5}, 35, 271

\bibitem[\protect\citeauthoryear{Wijchers, Dijkerman, Zeegers  \&
  Alkemade}{Wijchers et~al.}{1984}]{84WiDiZe.YO}
Wijchers T.,  Dijkerman H.~A.,  Zeegers P. J.~T.,   Alkemade C. T.~J.,  1984,
  \mn@doi [Chem. Phys.] {10.1016/0301-0104(84)80050-6}, 91, 141

\bibitem[\protect\citeauthoryear{Wyckoff \& Clegg}{Wyckoff \&
  Clegg}{1978}]{78wYcL.YO}
Wyckoff S.,  Clegg R. E.~S.,  1978, \mn@doi [MNRAS] {10.1093/mnras/184.1.127},
  184, 127

\bibitem[\protect\citeauthoryear{Yeo, Hummon, Collopy, Yan, Hemmerling, Chae,
  Doyle  \& Ye}{Yeo et~al.}{2015}]{15YeHuCo.YO}
Yeo M.,  Hummon M.~T.,  Collopy A.~L.,  Yan B.,  Hemmerling B.,  Chae E.,
  Doyle J.~M.,   Ye J.,  2015, \mn@doi [Phys. Rev. Lett.]
  {10.1103/PhysRevLett.114.223003}, 114, 223003

\bibitem[\protect\citeauthoryear{Yurchenko, Lodi, Tennyson  \&
  Stolyarov}{Yurchenko et~al.}{2016}]{Duo}
Yurchenko S.~N.,  Lodi L.,  Tennyson J.,   Stolyarov A.~V.,  2016, \mn@doi
  [Comput. Phys. Commun.] {10.1016/j.cpc.2015.12.021}, 202, 262

\bibitem[\protect\citeauthoryear{Yurchenko, Sinden, Lodi, Hill, Gorman  \&
  Tennyson}{Yurchenko et~al.}{2018a}]{jt711}
Yurchenko S.~N.,  Sinden F.,  Lodi L.,  Hill C.,  Gorman M.~N.,   Tennyson J.,
  2018a, \mn@doi [MNRAS] {10.1093/mnras/stx2738}, 473, 5324

\bibitem[\protect\citeauthoryear{{Yurchenko}, {Al-Refaie}  \&
  {Tennyson}}{{Yurchenko} et~al.}{2018b}]{ExoCross}
{Yurchenko} S.~N.,  {Al-Refaie} A.~F.,   {Tennyson} J.,  2018b, \mn@doi [A\&A]
  {10.1051/0004-6361/201732531}, 614, A131

\bibitem[\protect\citeauthoryear{Yurchenko, Nogu\'{e}, Azzam  \&
  Tennyson}{Yurchenko et~al.}{2022}]{22YuNoAz}
Yurchenko S.~N.,  Nogu\'{e} E.,  Azzam A. A.~A.,   Tennyson J.,  2022, \mn@doi
  [MNRAS] {10.1093/mnras/stac3757}, 520, 5183

\bibitem[\protect\citeauthoryear{Zhang, Zhang, Zhu, Gu, Suo, Chen  \&
  Zhao}{Zhang et~al.}{2017}]{17ZhZhZh.YO}
Zhang D.,  Zhang Q.,  Zhu B.,  Gu J.,  Suo B.,  Chen Y.,   Zhao D.,  2017,
  \mn@doi [J. Chem. Phys.] {10.1063/1.4978335}, 146, 114303

\makeatother
\end{thebibliography}






\bsp	
\label{lastpage}
\end{document}